\documentclass[%reprint,
superscriptaddress,
twocolumn,
preprintnumbers,
nofootinbib,
 amsmath,amssymb,
 aps,prd,
floatfix,
]{revtex4-2}

\usepackage{graphicx}% Include figure files
\usepackage{dcolumn}% Align table columns on decimal point
\usepackage{bm}% bold math
\usepackage[colorlinks,linkcolor=blue,citecolor=blue,urlcolor=blue ]{hyperref}% add hypertext capabilities
\usepackage{amsmath,amssymb,natbib,latexsym}
\usepackage[T1]{fontenc}
\usepackage[utf8]{inputenc}
\usepackage{physics}

\usepackage{hhline}% http://ctan.org/pkg/hhline
\usepackage{multirow}% http://ctan.org/pkg/multirow
\usepackage{xspace}
\usepackage{makecell}
\usepackage{ulem}

\usepackage{siunitx}

\usepackage{diagbox}

\newcommand{\camb}{\textsc{CAMB}\xspace}
\newcommand{\cosmolike}{\textsc{CosmoLike}\xspace}
\newcommand{\cosmoprimo}{\textsc{cosmoprimo}\xspace}
\newcommand{\cosmosis}{\textsc{CosmoSIS}\xspace}
\newcommand{\polychord}{\textsc{Polychord}\xspace}
\newcommand{\nautilus}{\textsc{nautilus}\xspace}
\newcommand{\getdist}{\textsc{GetDist}\xspace}
\newcommand{\halofit}{\textsc{Halofit}\xspace}

\newcommand{\healpix}{\textsc{HEALPix}\xspace}

\newcommand{\github}{{\tt GitHub}\xspace}
\newcommand{\baofitwtheta}{{\tt BAOfit\_wtheta}\xspace}
\newcommand{\tensiometer}{{\tt tensiometer}\xspace}

\usepackage{array}
\newcolumntype{Z}{>{\setbox0=\hbox\bgroup}c<{\egroup}@{}}

\usepackage{booktabs}

\usepackage{xcolor}

\newcommand{\zph}{z_\mathrm{ph}\xspace}

\newcommand{\mice}{{\tt MICE}\xspace}
\newcommand{\planck}{{\tt Planck}\xspace}

\newcommand{\lcdm}{\ensuremath{\Lambda{\rm CDM}}\xspace}
\newcommand{\klcdm}{\ensuremath{k\Lambda{\rm CDM}}\xspace}
\newcommand{\wcdm}{\ensuremath{w{\rm CDM}}\xspace}
\newcommand{\wacdm}{\ensuremath{w_0w_a{\rm CDM}}\xspace}
\newcommand{\nulcdm}{\ensuremath{\nu\Lambda{\rm CDM}}\xspace}

\newcommand{\ob}{\ensuremath{\Omega_\mathrm{b}}\xspace}
\newcommand{\om}{\ensuremath{\Omega_\mathrm{m}}\xspace}

\newcommand{\hubble}{\ensuremath{H_0}\xspace}
\newcommand{\kmsMpc}{km s$^{-1}$Mpc$^{-1}$}

\newcommand{\appendixcite}[1]{\hyperref[#1]{Appendix \ref*{#1}}}

\newcommand{\neutrinomass}{\ensuremath{\sum m_\nu}\xspace}
\newcommand{\sqdg}{${\rm deg}^2$\xspace}

\DeclareSIUnit\angstrom{\text {Å}}
\newcommand{\dec}{\ensuremath{\mathrm{Dec}}}

\newcommand{\tilesnodesi}{{\tt DR1tiles\_noDESI}\xspace}
\newcommand{\tilesonlydesi}{{\tt DR1tiles\_DESI}\xspace}
\newcommand{\decbelow}{{\tt Deccut\_noDESI}\xspace}
\newcommand{\decabove}{{\tt Deccut\_DESI}\xspace}

\begin{document}

\preprint{DES-2025-0903}
\preprint{FERMILAB-PUB-26-0028}

\title{Dark Energy Survey: DESI-Independent Angular BAO Measurement}

% Author list file generated with: mkauthlist 1.3.0+14.gcc6daf1.dirty 
% mkauthlist --sort --aux first_second_tier.csv DES-2025-0903_author_list.csv DES-2025-0903_author_list.tex 
%% Orcid numbers may need \usepackage{orcidlink}.
%% Use \input to call the file

\author{J. Mena-Fern{\'a}ndez}\email{menafernandez@cppm.in2p3.fr}
\affiliation{Universit\'e Grenoble Alpes, CNRS, LPSC-IN2P3, 38000 Grenoble, France}
\affiliation{Aix Marseille Univ, CNRS/IN2P3, CPPM, Marseille, France}

\author{S.~Avila}
\affiliation{Centro de Investigaciones Energ\'eticas, Medioambientales y Tecnol\'ogicas (CIEMAT), Madrid, Spain}

\author{A.~Porredon}
\affiliation{Centro de Investigaciones Energ\'eticas, Medioambientales y Tecnol\'ogicas (CIEMAT), Madrid, Spain}
\affiliation{Ruhr University Bochum, Faculty of Physics and Astronomy, Astronomical Institute, German Centre for Cosmological Lensing, 44780 Bochum, Germany}

\author{H.~Camacho}
\affiliation{Brookhaven National Laboratory, Bldg 510, Upton, NY 11973, USA}
\affiliation{Laborat\'orio Interinstitucional de e-Astronomia - LIneA, Av. Pastor Martin Luther King Jr, 126 Del Castilho, Nova Am\'erica Offices, Torre 3000/sala 817 CEP: 20765-000, Brazil}

\author{J.~Muir}
\affiliation{Department of Physics, University of Cincinnati, Cincinnati, Ohio 45221, USA}
\affiliation{Perimeter Institute for Theoretical Physics, 31 Caroline St. North, Waterloo, ON N2L 2Y5, Canada}

\author{E.~Sanchez}
\affiliation{Centro de Investigaciones Energ\'eticas, Medioambientales y Tecnol\'ogicas (CIEMAT), Madrid, Spain}

\author{M.~Adamow}
\affiliation{Center for Astrophysical Surveys, National Center for Supercomputing Applications, 1205 West Clark St., Urbana, IL 61801, USA}

\author{K.~Bechtol}
\affiliation{Physics Department, 2320 Chamberlin Hall, University of Wisconsin-Madison, 1150 University Avenue Madison, WI  53706-1390}

\author{R.~Camilleri}
\affiliation{School of Mathematics and Physics, University of Queensland,  Brisbane, QLD 4072, Australia}

\author{G.~Campailla}
\affiliation{Department of Physics, University of Genova and INFN, Via Dodecaneso 33, 16146, Genova, Italy}

\author{T.~M.~Davis}
\affiliation{School of Mathematics and Physics, University of Queensland,  Brisbane, QLD 4072, Australia}

\author{N.~Deiosso}
\affiliation{Centro de Investigaciones Energ\'eticas, Medioambientales y Tecnol\'ogicas (CIEMAT), Madrid, Spain}

\author{C.~Doux}
\affiliation{Department of Physics and Astronomy, University of Pennsylvania, Philadelphia, PA 19104, USA}
\affiliation{Universit\'e Grenoble Alpes, CNRS, LPSC-IN2P3, 38000 Grenoble, France}

\author{A.~Drlica-Wagner}
\affiliation{Department of Astronomy and Astrophysics, University of Chicago, Chicago, IL 60637, USA}
\affiliation{Fermi National Accelerator Laboratory, P. O. Box 500, Batavia, IL 60510, USA}
\affiliation{Kavli Institute for Cosmological Physics, University of Chicago, Chicago, IL 60637, USA}

\author{A.~Fert\'e}
\affiliation{SLAC National Accelerator Laboratory, Menlo Park, CA 94025, USA}

\author{R.~A.~Gruendl}
\affiliation{Center for Astrophysical Surveys, National Center for Supercomputing Applications, 1205 West Clark St., Urbana, IL 61801, USA}
\affiliation{Department of Astronomy, University of Illinois at Urbana-Champaign, 1002 W. Green Street, Urbana, IL 61801, USA}

\author{W.~G.~Hartley}
\affiliation{Department of Astronomy, University of Geneva, ch. d'\'Ecogia 16, CH-1290 Versoix, Switzerland}

\author{A.~Pieres}
\affiliation{Laborat\'orio Interinstitucional de e-Astronomia - LIneA, Av. Pastor Martin Luther King Jr, 126 Del Castilho, Nova Am\'erica Offices, Torre 3000/sala 817 CEP: 20765-000, Brazil}
\affiliation{Observat\'orio Nacional, Rua Gal. Jos\'e Cristino 77, Rio de Janeiro, RJ - 20921-400, Brazil}

\author{M.~Raveri}
\affiliation{Department of Physics, University of Genova and INFN, Via Dodecaneso 33, 16146, Genova, Italy}

\author{E.~S.~Rykoff}
\affiliation{Kavli Institute for Particle Astrophysics \& Cosmology, P. O. Box 2450, Stanford University, Stanford, CA 94305, USA}
\affiliation{SLAC National Accelerator Laboratory, Menlo Park, CA 94025, USA}

\author{I.~Sevilla-Noarbe}
\affiliation{Centro de Investigaciones Energ\'eticas, Medioambientales y Tecnol\'ogicas (CIEMAT), Madrid, Spain}

\author{P.~Shah}
\affiliation{Department of Physics \& Astronomy, University College London, Gower Street, London, WC1E 6BT, UK}

\author{E.~Sheldon}
\affiliation{Brookhaven National Laboratory, Bldg 510, Upton, NY 11973, USA}

\author{M.~Vincenzi}
\affiliation{Department of Physics, University of Oxford, Denys Wilkinson Building, Keble Road, Oxford OX1 3RH, United Kingdom}

\author{B.~Yanny}
\affiliation{Fermi National Accelerator Laboratory, P. O. Box 500, Batavia, IL 60510, USA}

\author{T.~M.~C.~Abbott}
\affiliation{Cerro Tololo Inter-American Observatory, NSF's National Optical-Infrared Astronomy Research Laboratory, Casilla 603, La Serena, Chile}

\author{M.~Aguena}
\affiliation{INAF-Osservatorio Astronomico di Trieste, via G. B. Tiepolo 11, I-34143 Trieste, Italy}
\affiliation{Laborat\'orio Interinstitucional de e-Astronomia - LIneA, Av. Pastor Martin Luther King Jr, 126 Del Castilho, Nova Am\'erica Offices, Torre 3000/sala 817 CEP: 20765-000, Brazil}

\author{S.~Allam}
\affiliation{Fermi National Accelerator Laboratory, P. O. Box 500, Batavia, IL 60510, USA}

\author{O.~Alves}
\affiliation{Department of Physics, University of Michigan, Ann Arbor, MI 48109, USA}

\author{F.~Andrade-Oliveira}
\affiliation{Physik-Institut, University of Zürich, Winterthurerstrasse 190, CH-8057 Zürich, Switzerland}

\author{J.~Annis}
\affiliation{Fermi National Accelerator Laboratory, P. O. Box 500, Batavia, IL 60510, USA}

\author{D.~Bacon}
\affiliation{Institute of Cosmology and Gravitation, University of Portsmouth, Portsmouth, PO1 3FX, UK}

\author{J.~Blazek}
\affiliation{Department of Physics, Northeastern University, Boston, MA 02115, USA}

\author{S.~Bocquet}
\affiliation{University Observatory, LMU Faculty of Physics, Scheinerstr. 1, 81679 Munich, Germany}

\author{D.~Brooks}
\affiliation{Department of Physics \& Astronomy, University College London, Gower Street, London, WC1E 6BT, UK}

\author{A.~Carnero~Rosell}
\affiliation{Instituto de Astrofisica de Canarias, E-38205 La Laguna, Tenerife, Spain}
\affiliation{Laborat\'orio Interinstitucional de e-Astronomia - LIneA, Av. Pastor Martin Luther King Jr, 126 Del Castilho, Nova Am\'erica Offices, Torre 3000/sala 817 CEP: 20765-000, Brazil}
\affiliation{Universidad de La Laguna, Dpto. Astrofísica, E-38206 La Laguna, Tenerife, Spain}

\author{J.~Carretero}
\affiliation{Institut de F\'{\i}sica d'Altes Energies (IFAE), The Barcelona Institute of Science and Technology, Campus UAB, 08193 Bellaterra (Barcelona) Spain}

\author{F.~J.~Castander}
\affiliation{Institut d'Estudis Espacials de Catalunya (IEEC), 08034 Barcelona, Spain}
\affiliation{Institute of Space Sciences (ICE, CSIC),  Campus UAB, Carrer de Can Magrans, s/n,  08193 Barcelona, Spain}

\author{R.~Cawthon}
\affiliation{Oxford College of Emory University, Oxford, GA 30054, USA}

\author{L.~N.~da Costa}
\affiliation{Laborat\'orio Interinstitucional de e-Astronomia - LIneA, Av. Pastor Martin Luther King Jr, 126 Del Castilho, Nova Am\'erica Offices, Torre 3000/sala 817 CEP: 20765-000, Brazil}

\author{M.~E.~da Silva Pereira}
\affiliation{Hamburger Sternwarte, Universit\"{a}t Hamburg, Gojenbergsweg 112, 21029 Hamburg, Germany}

\author{J.~De~Vicente}
\affiliation{Centro de Investigaciones Energ\'eticas, Medioambientales y Tecnol\'ogicas (CIEMAT), Madrid, Spain}

\author{S.~Desai}
\affiliation{Department of Physics, IIT Hyderabad, Kandi, Telangana 502285, India}

\author{H.~T.~Diehl}
\affiliation{Fermi National Accelerator Laboratory, P. O. Box 500, Batavia, IL 60510, USA}

\author{B.~Flaugher}
\affiliation{Fermi National Accelerator Laboratory, P. O. Box 500, Batavia, IL 60510, USA}

\author{J.~Frieman}
\affiliation{Department of Astronomy and Astrophysics, University of Chicago, Chicago, IL 60637, USA}
\affiliation{Fermi National Accelerator Laboratory, P. O. Box 500, Batavia, IL 60510, USA}
\affiliation{Kavli Institute for Cosmological Physics, University of Chicago, Chicago, IL 60637, USA}

\author{J.~Garc\'ia-Bellido}
\affiliation{Instituto de Fisica Teorica UAM/CSIC, Universidad Autonoma de Madrid, 28049 Madrid, Spain}

\author{M.~Gatti}
\affiliation{Kavli Institute for Cosmological Physics, University of Chicago, Chicago, IL 60637, USA}

\author{D.~Gruen}
\affiliation{University Observatory, LMU Faculty of Physics, Scheinerstr. 1, 81679 Munich, Germany}

\author{G.~Gutierrez}
\affiliation{Fermi National Accelerator Laboratory, P. O. Box 500, Batavia, IL 60510, USA}

\author{K.~Herner}
\affiliation{Fermi National Accelerator Laboratory, P. O. Box 500, Batavia, IL 60510, USA}

\author{S.~R.~Hinton}
\affiliation{School of Mathematics and Physics, University of Queensland,  Brisbane, QLD 4072, Australia}

\author{D.~L.~Hollowood}
\affiliation{Santa Cruz Institute for Particle Physics, Santa Cruz, CA 95064, USA}

\author{K.~Honscheid}
\affiliation{Center for Cosmology and Astro-Particle Physics, The Ohio State University, Columbus, OH 43210, USA}
\affiliation{Department of Physics, The Ohio State University, Columbus, OH 43210, USA}

\author{D.~Huterer}
\affiliation{Department of Physics, University of Michigan, Ann Arbor, MI 48109, USA}

\author{N.~Jeffrey}
\affiliation{Department of Physics \& Astronomy, University College London, Gower Street, London, WC1E 6BT, UK}

\author{K.~Kuehn}
\affiliation{Australian Astronomical Optics, Macquarie University, North Ryde, NSW 2113, Australia}
\affiliation{Lowell Observatory, 1400 Mars Hill Rd, Flagstaff, AZ 86001, USA}

\author{O.~Lahav}
\affiliation{Department of Physics \& Astronomy, University College London, Gower Street, London, WC1E 6BT, UK}

\author{S.~Lee}
\affiliation{Jet Propulsion Laboratory, California Institute of Technology, 4800 Oak Grove Dr., Pasadena, CA 91109, USA}

\author{J.~L.~Marshall}
\affiliation{George P. and Cynthia Woods Mitchell Institute for Fundamental Physics and Astronomy, and Department of Physics and Astronomy, Texas A\&M University, College Station, TX 77843,  USA}

\author{F.~Menanteau}
\affiliation{Center for Astrophysical Surveys, National Center for Supercomputing Applications, 1205 West Clark St., Urbana, IL 61801, USA}
\affiliation{Department of Astronomy, University of Illinois at Urbana-Champaign, 1002 W. Green Street, Urbana, IL 61801, USA}

\author{R.~Miquel}
\affiliation{Instituci\'o Catalana de Recerca i Estudis Avan\c{c}ats, E-08010 Barcelona, Spain}
\affiliation{Institut de F\'{\i}sica d'Altes Energies (IFAE), The Barcelona Institute of Science and Technology, Campus UAB, 08193 Bellaterra (Barcelona) Spain}

\author{J.~J.~Mohr}
\affiliation{University Observatory, LMU Faculty of Physics, Scheinerstr. 1, 81679 Munich, Germany}

\author{J.~Myles}
\affiliation{Department of Astrophysical Sciences, Princeton University, Peyton Hall, Princeton, NJ 08544, USA}

\author{R.~L.~C.~Ogando}
\affiliation{Centro de Tecnologia da Informa\c{c}\~ao Renato Archer, Campinas, SP, Brazil - 13069-901}
\affiliation{Observat\'orio Nacional, Rio de Janeiro, RJ, Brazil - 20921-400}

\author{A.~Palmese}
\affiliation{Department of Physics, Carnegie Mellon University, Pittsburgh, Pennsylvania 15312, USA}

\author{W.~J.~Percival}
\affiliation{Department of Physics and Astronomy, University of Waterloo, 200 University Ave W, Waterloo, ON N2L 3G1, Canada}
\affiliation{Perimeter Institute for Theoretical Physics, 31 Caroline St. North, Waterloo, ON N2L 2Y5, Canada}

\author{A.~A.~Plazas~Malag\'on}
\affiliation{Kavli Institute for Particle Astrophysics \& Cosmology, P. O. Box 2450, Stanford University, Stanford, CA 94305, USA}
\affiliation{SLAC National Accelerator Laboratory, Menlo Park, CA 94025, USA}

\author{M.~Rodriguez-Monroy}
\affiliation{Ruhr University Bochum, Faculty of Physics and Astronomy, Astronomical Institute, German Centre for Cosmological Lensing, 44780 Bochum, Germany}

\author{A.~Roodman}
\affiliation{Kavli Institute for Particle Astrophysics \& Cosmology, P. O. Box 2450, Stanford University, Stanford, CA 94305, USA}
\affiliation{SLAC National Accelerator Laboratory, Menlo Park, CA 94025, USA}

\author{S.~Samuroff}
\affiliation{Department of Physics, Northeastern University, Boston, MA 02115, USA}
\affiliation{Institut de F\'{\i}sica d'Altes Energies (IFAE), The Barcelona Institute of Science and Technology, Campus UAB, 08193 Bellaterra (Barcelona) Spain}

\author{D.~Sanchez Cid}
\affiliation{Centro de Investigaciones Energ\'eticas, Medioambientales y Tecnol\'ogicas (CIEMAT), Madrid, Spain}
\affiliation{Physik-Institut, University of Zürich, Winterthurerstrasse 190, CH-8057 Zürich, Switzerland}

\author{B.~O.~S\'anchez}
\affiliation{Aix Marseille Univ, CNRS/IN2P3, CPPM, Marseille, France}

\author{T.~Shin}
\affiliation{Department of Physics and Astronomy, Stony Brook University, Stony Brook, NY 11794, USA}

\author{M.~Smith}
\affiliation{Physics Department, Lancaster University, Lancaster, LA1 4YB, UK}

\author{M.~Soares-Santos}
\affiliation{Physik-Institut, University of Zürich, Winterthurerstrasse 190, CH-8057 Zürich, Switzerland}

\author{E.~Suchyta}
\affiliation{Computer Science and Mathematics Division, Oak Ridge National Laboratory, Oak Ridge, TN 37831}

\author{M.~E.~C.~Swanson}
\affiliation{Center for Astrophysical Surveys, National Center for Supercomputing Applications, 1205 West Clark St., Urbana, IL 61801, USA}

\author{D.~L.~Tucker}
\affiliation{Fermi National Accelerator Laboratory, P. O. Box 500, Batavia, IL 60510, USA}

\author{V.~Vikram}
\affiliation{Department of Physics, Central University of Kerala, Kasaragod, Kerala, India}

\author{A.~R.~Walker}
\affiliation{Cerro Tololo Inter-American Observatory, NSF's National Optical-Infrared Astronomy Research Laboratory, Casilla 603, La Serena, Chile}

\author{N.~Weaverdyck}
\affiliation{Berkeley Center for Cosmological Physics, Department of Physics, University of California, Berkeley, CA 94720, US}
\affiliation{Lawrence Berkeley National Laboratory, 1 Cyclotron Road, Berkeley, CA 94720, USA}

\author{M.~Yamamoto}
\affiliation{Department of Astrophysical Sciences, Princeton University, Peyton Hall, Princeton, NJ 08544, USA}
\affiliation{Department of Physics, Duke University Durham, NC 27708, USA}

\collaboration{DES Collaboration}

\date{\today}% It is always \today, today,

\begin{abstract}
    We present a measurement of the angular baryon acoustic oscillation (BAO) scale from the completed Dark Energy Survey (DES) dataset excluding the area of overlap with the Dark Energy Spectroscopic Instrument (DESI). We follow the same methodology and validation process as in the DES year 6 (Y6) BAO analysis. We interpret the impact of this measurement in the context of the statistical preference for $w_0w_a$ cold dark matter (CDM) over $\Lambda$CDM when combined with DES Y5 Type Ia supernovae (SN), Planck Cosmic Microwave Background (CMB) and DESI BAO. Based on our previous work, using the full Y6 DES BAO sample, in combination with SN, CMB and DESI data release 1 (DR1) BAO, added $0.3\sigma$ in this preference (from $3.7\sigma$ to $4.0\sigma$), but this ignored possible correlations between datasets. Using our new DESI-independent DES BAO likelihood instead, we find a smaller increase in the statistical preference for $w_0w_a$CDM, from $3.7\sigma$ to $3.8\sigma$ when using DESI DR1 BAO, and from $4.0\sigma$ to $4.1\sigma$ when updating to the more recent DESI data release 2 (DR2) BAO. These significances reduce to $3.1\sigma$ when using the new calibrated DES SN-Dovekie. Alongside this work, we publicly release \texttt{BAOfit\_wtheta}, the BAO fitting code for the angular correlation function used in the DES Y6 BAO analysis.

\end{abstract}

\maketitle

%\tableofcontents

\section{\label{sec:intro}Introduction}

Baryon acoustic oscillations (BAO) provide one of the most robust and well-understood probes of the expansion history of the Universe. Originating from sound waves propagating in the early Universe, the BAO feature leaves a characteristic imprint in the large-scale distribution of matter, visible today in the clustering of galaxies and the Cosmic Microwave Background (CMB). Since its first detections in galaxy redshift surveys \cite{eisenstein2005detection,cole20052df}, the BAO signal has become a cornerstone of precision cosmology, offering a ``standard ruler'' to measure cosmological distances and constrain the nature of dark energy.

Over the past two decades, spectroscopic surveys such as the 6-degree Field Galaxy Survey (6dFGS) \citep{beutler20116df}, the WiggleZ dark energy survey \citep{blake2011wigglez_1,blake2011wigglez_2,kazin2014wigglez,hinton2016measuring}, and the [extended] Baryonic Oscillations Spectroscopic Survey ([e]BOSS), part of the Sloan Digital Sky Survey (SDSS) series \cite{ross2015clustering,alam2017clustering,ata2018clustering,delubac2013baryon,slosar2013measurement,font2014quasar, delubac2015baryon,bautista2017measurement,des2020completed,alam2021completed}, have provided increasingly precise BAO measurements across a wide range of redshifts. Building on this legacy, the Dark Energy Spectroscopic Instrument (DESI) is now delivering even more precise and higher-redshift BAO measurements \cite{desi2024iii,desidr2bao}. Complementary to these, photometric surveys enable BAO studies over much larger volumes, albeit with reduced radial resolution. Recent analyses from the Dark Energy Survey (DES) \cite{abbott2019dark,abbott2022dark,y6-baokp} and the Dark
Energy Camera Legacy Survey (DECaLS) \cite{song2024measurement,saulder2025studying} have demonstrated the feasibility and competitiveness of photometric BAO measurements, reinforcing their role as a cross-check on spectroscopic results.

DES was designed to probe the origin of cosmic acceleration by mapping hundreds of millions of galaxies and measuring the large-scale structure (LSS) of the Universe with unprecedented precision \cite{dark2005dark}. Conducted with the 570-megapixel Dark Energy Camera (DECam) mounted on the 4-meter Blanco Telescope at Cerro Tololo Inter-American Observatory in Chile \cite{flaugher2015dark}, DES observed approximately 5,000 \sqdg of the southern sky in five optical–near-infrared bands ($grizY$) over six observing seasons from 2013 to 2019. DES uses several complementary cosmological probes, including, but not limited to, weak gravitational lensing, galaxy clustering (including BAO), Type Ia supernovae and galaxy cluster abundances. Importantly for BAO science, the survey’s depth and area enable precise angular galaxy clustering measurements out to a redshift of about 1.2, providing one of the largest and most homogeneous photometric datasets to date.

Using the complete DES dataset, covering 6 years of DES observations (DES Y6), we measured the angular BAO scale at an effective redshift of $z_\mathrm{eff}=0.851$ \cite{y6-baokp}. We used a galaxy sample optimized for BAO science in the redshift range $0.6<z<1.2$, comprising nearly 16 million galaxies over 4,273.42 \sqdg \cite{mena2024dark}, hereafter the DES Y6 BAO sample. To mitigate confirmation bias, the analysis was performed blind. A battery of pre-unblinding tests had to be passed before unblinding the final result, together with many robustness checks. The consensus measurement constrained the ratio of the angular distance to the sound horizon scale to $D_M(z_\mathrm{eff})/r_d=19.51\pm0.41$, representing a 2.1\% precision, marking it as the most precise angular BAO measurement from any photometric survey. Notably, this measurement was found to be 4.3\% and $2.1\sigma$ below the angular BAO scale predicted by Planck when assuming the $\Lambda$ cold dark matter (CDM) model.

On the other hand, DESI provides the state-of-the-art three-dimensional BAO measurements. DESI maps the LSS of the Universe by obtaining optical spectra for tens of millions of galaxies and quasars, constructing a 3D map spanning the nearby Universe out to redshift of about 3.5 \cite{aghamousa2016desi_1,aghamousa2016desi_2}. The DESI survey footprint encompasses approximately 14,000 \sqdg, covering regions in both the north and south galactic caps, with a small overlap with DES in the southern hemisphere. One of DESI’s primary scientific goals is to measure the BAO signal with unprecedented precision, both along and across the line of sight. The BAO measurements and cosmological constraints using the DESI Data Release 1 (DR1; first year of DESI data) were released in 2024 \cite{desi2024iii,desi2024iv}, and the ones for the Data Release 2 (DR2; first three years of DESI data), in 2025 \cite{desidr2bao}.

In a recent study \cite{abbott2025dark}, we investigated the cosmological implications of a joint analysis of the two DES expansion history probes—BAO \cite{y6-baokp} and Type Ia supernovae (SN) \cite{des-y5-sn}—combined with external datasets, namely the CMB from Planck \cite{aghanim2020planckVI}, within the framework of \lcdm and several extended models: \klcdm, \wcdm, \wacdm, and \nulcdm. We found tensions among the datasets in \lcdm, \klcdm, \wcdm, and \nulcdm, indicating that no combination of parameters within these models could simultaneously explain all observations. The \wacdm model, which allows a time-dependent dark energy equation of state, was the only one capable of reconciling the different observables, and therefore we focus on it in what follows.

Combining DES BAO, DES SN and CMB, we found a $3.0\sigma$ preference for \wacdm according to the $\Delta\chi^2$ criterion. By contrast, combining DES SN, CMB and DESI DR1 BAO gave $3.7\sigma$, which further increased to $4.0\sigma$ when DES BAO was also included. This $0.3\sigma$ increase is the primary motivation for the analysis presented in this paper. However, because the DES and DESI survey footprints partially overlap, a direct combination of their BAO constraints is non-trivial, as correlations between the two datasets are expected—an effect that was neglected in \cite{abbott2025dark}. Since this work builds upon the analysis of \cite{abbott2025dark}, which relied on DESI DR1 BAO data, here we include both the DESI DR1 and the more recent DR2 BAO measurements.

The first part of this analysis is devoted to constructing a DES BAO likelihood that can be directly combined with that of DESI. To achieve this, we exclude the region of the sky where the DES and DESI samples overlap, and measure the BAO signal in the remaining DES footprint. We refer to it as ``DES BAO-noDESI''. This procedure leads to slightly larger uncertainties compared to our fiducial analysis (i.e., compared to including all the DES area), but ensures that the resulting measurements are statistically independent of those from DESI. We construct two alternative DES BAO likelihoods depending on how much DES area we exclude: one designed for combination with DESI DR1/2, which forms the basis of the cosmological analysis in this work, and another intended for future joint analyses with DESI DR3/4. The robustness of our BAO measurements is further established through validation against 1,952 galaxy mock catalogs, designed to match the properties of the DES data.

The second part of this work focuses on deriving constraints on cosmological parameters within both \lcdm and its extension \wacdm. We combine the DES Y5 SN sample \cite{des-y5-sn}—used as our baseline in this analysis for consistency with \cite{abbott2025dark}—with Planck CMB data \cite{aghanim2020planckVI}, DESI DR1/2 BAO measurements \cite{desi2024iii,desi2024iv,desidr2bao}, and the DES BAO-noDESI measurement presented in this work. The combination of all these datasets allows us to improve constraints on \wacdm and assess whether the resulting level of tension with \lcdm is strengthened or alleviated. Finally, we also examine how adopting the newly calibrated DES SN–Dovekie sample \cite{popovic2025dark} affects our cosmological constraints and the inferred tension with \lcdm.

The structure of the paper is as follows: in \autoref{sec:bao_analysis}, we present the methodology followed to exclude the DESI area from the DES Y6 BAO sample, which we also apply to the simulations; in \autoref{sec:bao_methodology} we summarize the methodology followed to measure the BAO feature on the two-point angular correlation function, including the estimation of the covariances and the sources of BAO systematics; in \autoref{sec:bao_measurements}, we present the BAO measurements on our data samples; in \autoref{sec:cosmo_constraints}, we describe the inference of cosmological parameters and the results in both \lcdm and \wacdm models; and in \autoref{sec:conclusions}, we present the conclusions to our analysis.

\section{Data and simulations}\label{sec:bao_analysis}

One of the main goals of this paper is to provide an alternative DES Y6 BAO likelihood that is fully independent of DESI. To achieve this, we remove the region of sky where DES and DESI overlap, ensuring that the two samples are statistically independent. 

In this section, we describe the DES Y6 BAO sample and simulations, together with the procedure to remove their overlap with DESI. All the data products of the DES Y6 BAO analysis we use here (sample, simulations, mask) are publicly available at the DES database.\footnote{\url{https://des.ncsa.illinois.edu/releases/y6a2/Y6bao}.}

\subsection{The DESI-independent DES BAO sample}

\subsubsection{The DES Y6 BAO sample}\label{sec:des_y6_sample}

The DES Y6 BAO sample is described in detail in \citet{mena2024dark}. This sample was created using a red selection given by
\begin{equation}
    1.7< (i-z) + 2(r-i),
    \label{eq:redselection}
\end{equation}
since red galaxies are expected to have better redshift estimates and higher galaxy bias, both of which improve the expected BAO signal. It also includes a cut in the $i$ band
\begin{equation}\label{eq:fluxcut}
    i<22.5
\end{equation}
to ensure high quality in our validation pipeline, and a redshift-dependent magnitude limit, also in the $i$ band, given by
\begin{equation}\label{eq:maglim}
    i<19.64+2.894z_\mathrm{ph},
\end{equation}
which was re-optimized in \cite{mena2024dark} for the DES Y6 analysis to maximize the BAO signal. Its redshift range is 
\begin{equation}\label{eq:redshiftrange}
    0.6<\zph<1.2,
\end{equation} 
where $\zph$ is the photometric redshift estimate (see \cite{y6-baokp} for further details), and its effective redshift is $z_\mathrm{eff} = 0.851$. For most of the analysis, we divide it into 6 tomographic bins of width $\Delta \zph=0.1$.

\subsubsection{Excluding the DESI area}\label{sec:mask}

\begin{figure*}
    \centering
    \includegraphics[width=0.49\linewidth]{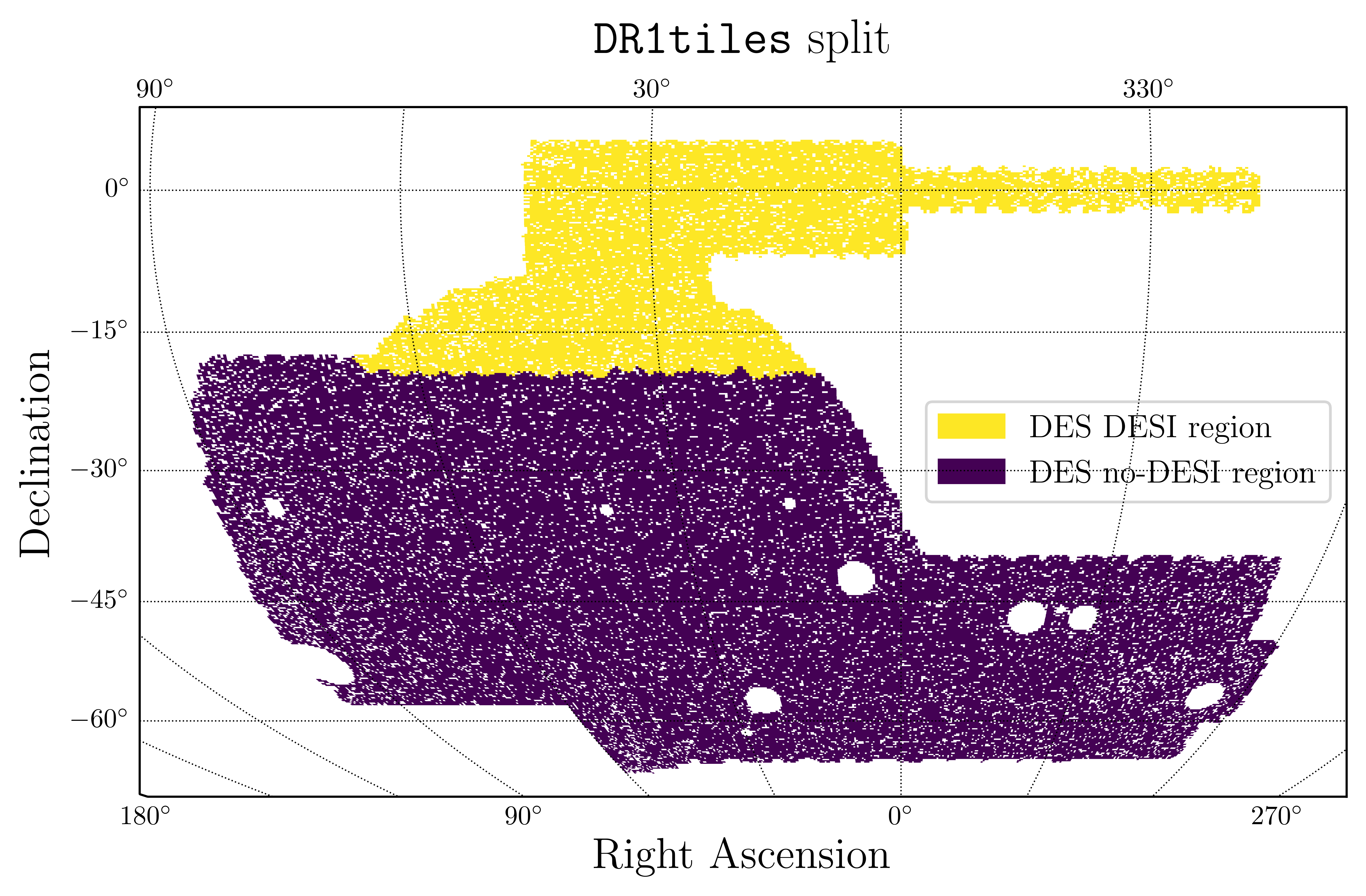}
    \hfill
    \includegraphics[width=0.49\linewidth]{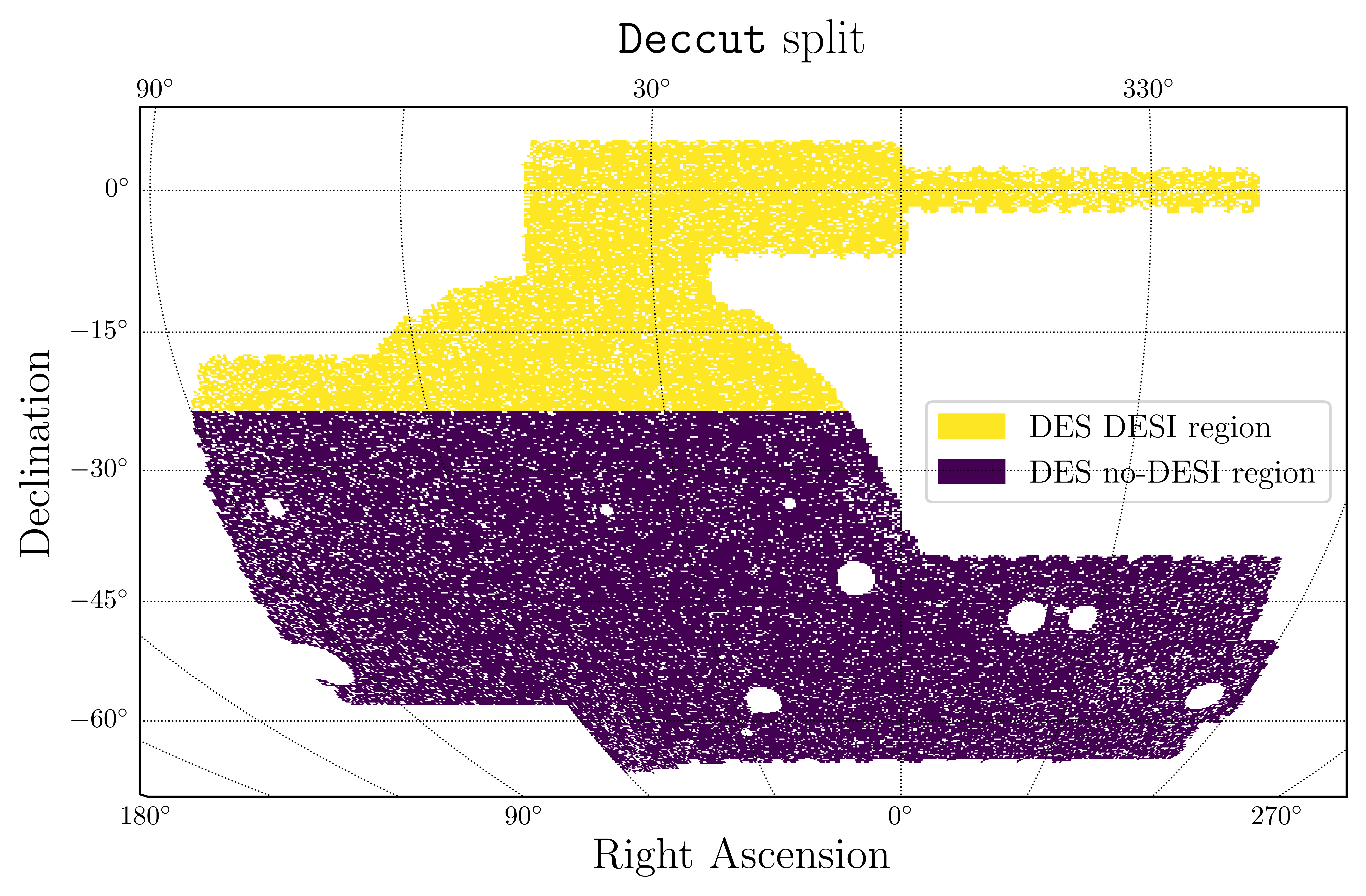}
    \caption{Splits of the DES Y6 BAO angular mask into the overlapping area with DESI (yellow, DES DESI region) and the non-overlapping one (purple, DES no-DESI region). \textbf{Left:} split performed using a DESI mask created from the DESI DR1 tiles. \textbf{Right:} split performed with a cut at $\dec = -23.5$ deg.}
    \label{fig:mask_splits}
\end{figure*}

We begin with the same angular mask used in the fiducial DES Y6 BAO analysis, covering an area of 4,273.42 \sqdg and described in \cite{mena2024dark}. Our objective is to divide this mask into two complementary parts: one excluding the DESI footprint (the DES no-DESI region) and one retaining only the overlapping area (the DES DESI region). We consider two approaches for this split:  
\begin{itemize}  
    \item \textbf{\tt DR1tiles} split. Designed for combination with DESI DR1 or DR2. In this case, we use the DESI angular mask constructed from the publicly available DESI DR1 tiles\footnote{\url{https://data.desi.lbl.gov/public/dr1/survey/ops/surveyops/tags/1.0/ops/tiles-main.ecsv}.} to exclude its overlap with the DES Y6 BAO mask. The resulting no-DESI mask covers 3,224.72 \sqdg, which we denote as \tilesnodesi. The complementary DES DESI mask spans 1,048.72 \sqdg and is denoted as \tilesonlydesi. Both masks are shown in the left panel of \autoref{fig:mask_splits}.  

    \item \textbf{\tt Deccut} split. Designed for future combinations with DESI DR3 and DR4. Here, we retain only pixels with $\dec < -23.5$ deg in the DES mask, following the DESI DR4 survey plan to extend to this declination. The resulting no-DESI mask has an area of 2,935.23 \sqdg, referred to as \decbelow, while the complementary DES DESI mask covers 1,338.19 \sqdg and is referred to as \decabove. These masks are shown in the right panel of \autoref{fig:mask_splits}.  
\end{itemize}

The sample used for the combination with DESI DR1/2 (\tilesnodesi) covers a slightly larger area than the sample designed for the combination with DR3/4 (\decbelow). As a result, the statistical uncertainty of the BAO measurement is expected to be smaller for the former. In this work, we therefore adopt the DR2-compatible sample as our default, as DR2 is the most recent DESI BAO dataset available at the time of this analysis. The complementary DES DESI samples (\tilesonlydesi and \decabove) are used to perform consistency checks.

The effective redshift of the different splits is the same as that of the full sample, $z_\mathrm{eff}=0.851$. The redshift distributions are the same compared to the full sample, and therefore we use the same $n(z)$ as in the fiducial Y6 BAO analysis for both the theoretical templates and the covariances (see \cite{mena2024dark}). The different masks are summarized in \autoref{tab:areas_masks}, where we include their descriptions and covered areas. In \autoref{tab:ngal_samples}, we display the number of galaxies per redshift bin of the different samples after applying each of the masks and the full sample for comparison purposes.

\begin{table*}[]
    \renewcommand{\arraystretch}{1.3} % Default value: 1
    \setlength{\tabcolsep}{4pt} % Adjust column separation
    \centering
    \begin{tabular}{l|l|l|c}
        \toprule\toprule
        Name & Description & Goal & Area (\sqdg) \\\hline
        Full mask & Fiducial DES Y6 BAO mask & -- & 4,273.42 \\\hline
        \tilesnodesi & \makecell[l]{Removed the overlap with DESI \\ using mask from public DR1 tiles} & Combination with DESI DR1/2 & 3,224.72 \\
        \tilesonlydesi & Complementary to \tilesnodesi & Cross-check & 1,048.72 \\\hline
        \decbelow & \makecell[l]{Kept $\dec<-23.5$} & \makecell[l]{Combination with DESI DR3/4} & 2,935.23\\
        \decabove & Complementary to \decbelow & Cross-check & 1,338.19 \\
        \bottomrule\bottomrule
    \end{tabular}
    \caption{Different masks used in this paper and the area they cover in squared degrees. The area of the \tilesnodesi sample, which we use for cosmological inference in this analysis, represents 75.5\% of the full footprint.}
    \label{tab:areas_masks}
\end{table*}

\begin{table*}[]
    \renewcommand{\arraystretch}{1.3} % Default value: 1
    \setlength{\tabcolsep}{4pt} % Adjust column separation
    \centering
    \begin{tabular}{l|cccccc|c}
        \toprule\toprule
        \multirow{2}{*}{\diagbox{Sample}{$z_{\rm ph}$ bin}} & Bin 1 & Bin 2 & Bin 3 & Bin 4 & Bin 5 & Bin 6 & Total \\
        & [0.6, 0.7] & [0.7, 0.8] & [0.8, 0.9] & [0.9, 1.0] & [1.0, 1.1] & [1.1, 1.2] & [0.6, 1.2] \\\hline
        Full sample    & 2,854,542 & 3,266,097 & 3,898,672 & 3,404,744 & 1,752,169 & 761,332 & 15,937,556 \\\hline
        \tilesnodesi   & 2,162,619 & 2,472,245 & 2,949,553 & 2,562,757 & 1,315,824 & 571,741 & 12,034,739 \\
        \tilesonlydesi & 691,923 & 793,852 & 949,119 & 841,987 & 436,345 & 189,591 & 3,902,817 \\\hline
        \decbelow      & 1,965,506 & 2,252,526 & 2,695,701 & 2,333,589 & 1,194,901 & 519,579 & 10,961,802 \\
        \decabove      & 889,036 & 1,013,571 & 1,202,971 & 1,071,155 & 557,268 & 241,753 & 4,975,754 \\
        \bottomrule\bottomrule
    \end{tabular}
    \caption{Number of galaxies per redshift bin for each of the BAO sample splits and the full sample for comparison purposes.}
    \label{tab:ngal_samples}
\end{table*}

\subsubsection{Mitigating LSS systematics}\label{sec:lss_sys}

The procedure used to mitigate imaging systematics in the clustering signal for the DES Y6 BAO analysis is described in \cite{mena2024dark,y6-baokp}, where the Iterative Systematic Decontamination (ISD) algorithm was applied \cite{crocce2016galaxy,elvin2018dark,rosell2022dark,rodriguez2022dark,rodriguez2025dark}. In short, ISD identifies correlations between galaxy density and a set of imaging properties, iteratively removes them and produces a \healpix weight map that is applied when measuring the two-point correlation functions (see \autoref{sec:clustering_statistics}). A more thorough description of the weights for the BAO sample can be found in \cite{mena2024dark}.

\subsection{Clustering statistics}\label{sec:clustering_statistics}

Here, we describe the production of the random catalogs and the measurements of the two-point angular correlation functions that we use as data vectors to measure the BAO signal.

The starting point to measure all clustering statistics is the creation of a random catalog. We create an independent random catalog for each of the 4 galaxy samples considered in this paper, each of them having 20 times as many objects as its corresponding sample. This amount of randoms was found to be more than sufficient to avoid any additional noise coming from them, as discussed in \cite{y6-baokp}. The random catalog is created by sampling the masks described in \autoref{sec:mask} while downsampling the pixels according to their coverage fraction, which is always larger than 80\%. 

Once we have the random sample, the two-point angular correlation function (ACF) is estimated using the Landy-Szalay estimator \cite{landy1993bias} as
\begin{equation}
    w(\theta) = \frac{DD(\theta)-2DR(\theta)+RR(\theta)}{RR(\theta)},
\end{equation}
where $DD$, $DR$ and $RR$ are the normalized counts of data-data, data-random and random-random pairs, respectively, separated by $\theta\pm\Delta\theta/2$, with $\Delta\theta$ being the bin size. We measure the ACF with a bin size of $\Delta\theta = $0.05 deg, but we also combine them in broader bins (in particular, $\Delta\theta = 0.10$, 0.15, 0.20 and 0.25 deg). As in \cite{y6-baokp}, our default binning is set to $\Delta\theta = 0.20$ deg.

\begin{figure*}
    \centering
    \includegraphics[width=\linewidth]{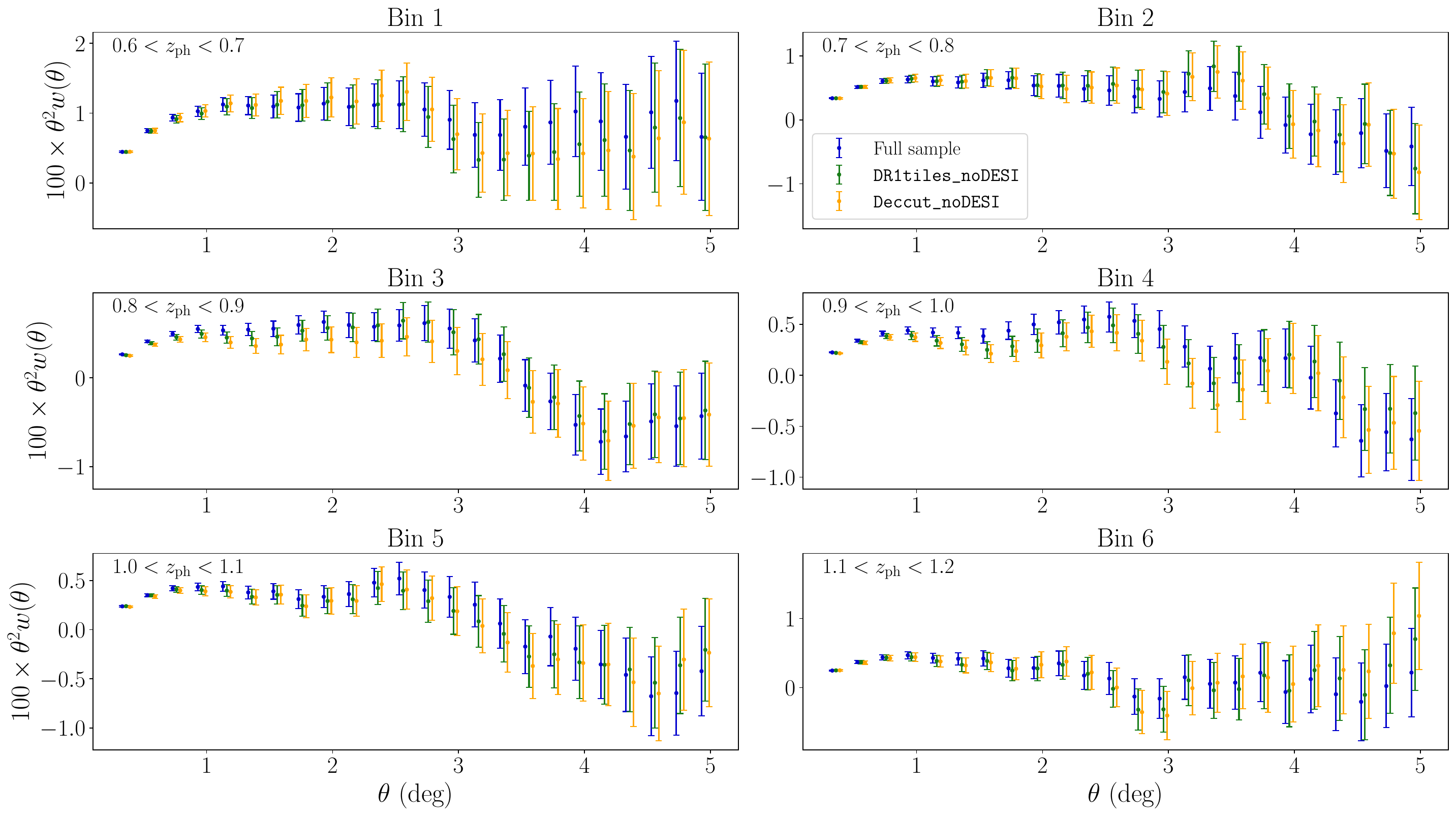}
    \caption{Angular correlation functions, or $w(\theta)$, for the full sample and the two no-DESI samples (\tilesnodesi and \decbelow). The error bars come from the \cosmolike covariance matrices, described in \autoref{sec:cov}, computed for each of the samples. Points are slightly shifted horizontally to avoid overlap.}
    \label{fig:wtheta_samples}
\end{figure*}

The angular correlation functions for the two no-DESI samples are shown in \autoref{fig:wtheta_samples}, together with those of the full sample for comparison purposes. The error bars come from the \cosmolike covariance matrices, described in \autoref{sec:cov}, computed for each of the samples. We see that the three of them are consistent.

\subsection{The COLA mocks}

We use the same mock catalogs as in the fiducial DES Y6 BAO analysis \cite{y6-baokp} for the validation of our BAO analysis. These mocks were produced using the \texttt{ICE-COLA} code \cite{izard2016ice}, a lightcone implementation of \texttt{COLA} \cite{koda2016fast} that combines second-order Lagrangian Perturbation Theory with a particle-mesh gravity solver, allowing us to efficiently simulate large cosmological volumes. Galaxies were populated using a hybrid Halo Occupation Distribution and Halo Abundance Matching model, with two free parameters per tomographic bin.

We have a total of 1,952 COLA mocks designed to closely match several key properties of the data, such as the observed volume, galaxy number density, true and photometric redshift distributions, and the evolution of clustering with redshift. Each simulation contains $2048^3$ particles in a cubic box of length 1536~$\mathrm{Mpc}/h$. We replicate this volume 64 times using periodic boundary conditions to cover the full sky and boost the statistical power without dramatically increasing computational cost, which gives us a lightcone up to $z=1.43$.
The production and validation are described in \cite{y6-baokp}, following the methodology from \cite{ferrero2021dark}, which builds up on \cite{avila2018dark}. Hereafter, we refer to them as COLA mocks. 

The underlying cosmological parameters of the simulations match those of the \textsc{MICE} \textit{Grand Challenge} simulation \cite{fosalba2015mice,crocce2015mice}, hereafter denoted as the \mice cosmology.
% , and we will use it as the cosmology for the template when running the BAO fits on the mocks. 
The \mice cosmological parameters are explicitly listed in \autoref{tab:cosmo_params}.

We apply the masks listed in \autoref{tab:areas_masks} to the 1,952 COLA mocks and measure the angular two-point correlation functions for all of them ($4\ {\rm masks}\times 1952\ {\rm mocks}\times 6\ {\rm redshift\ bins}=46,848\ w(\theta)$ measurements), following the methodology described in \autoref{sec:clustering_statistics}, to perform a one-on-one comparison with the analysis of the data. 
% The results are shown in \autoref{sec:mocks_results}.
We note that these mocks are not used for computing covariances (see \autoref{sec:cov}) due to the replications of the boxes explained above, which were found to yield spurious correlations in \cite{ferrero2021dark}. The main goal of these mocks is to validate the analysis and interpret statistical fluctuations.

\begin{table}[]
    \renewcommand{\arraystretch}{1.3} % Default value: 1
    \setlength{\tabcolsep}{4pt} % Adjust column separation
    \centering
    \begin{tabular}{c|cc}
        \toprule\toprule
        & \mice & \planck\\\hline
        \om & 0.25 & 0.31 \\
        \ob & 0.044 & 0.0481 \\
        $h$ & 0.7 & 0.676 \\
        $\sigma_8$ & 0.8 & 0.8 \\
        $n_s$ & 0.95 & 0.97 \\
        % Dataset & COLA mocks & Data \\
        \bottomrule\bottomrule
    \end{tabular}
    \caption{Cosmological parameters used for the template when measuring the BAO feature on the COLA mocks (\mice cosmology) and the DES BAO data (\planck cosmology).}
    \label{tab:cosmo_params}
\end{table}

\section{BAO methodology}\label{sec:bao_methodology}

In this section, we present the methodology adopted to measure the BAO signal from the two-point angular correlation function, describe the covariance matrices used in the BAO fits, and discuss the treatment of systematic uncertainties affecting the measurement. We closely follow the same methodology as in the fiducial DES Y6 BAO analysis, see \cite{y6-baokp}, but also highlight the main differences.

\subsection{The BAO fit}

We extract the BAO feature from the two-point angular correlation function using a template-based method. The methodology followed is described in detail in \cite{y6-baokp} and \cite{y3-baokp}, and is based on \cite{chan2018bao}. The specific implementation used here is released together with this paper in  \baofitwtheta,\footnote{\url{https://github.com/juanejo95/BAOfit_wtheta}.} and it is a minor update with respect to the code used for the fiducial angular correlation function analysis of \cite{y6-baokp}. A detailed description of the code’s functionality can be found in \appendixcite{app:baofitwtheta}. We summarize the methodology below.   

We start from the linear matter power spectrum generated using \camb \cite{lewis2000efficient}. We include the broadening of the BAO peak due to non-linearities \cite{crocce2008nonlinear,padmanabhan2009calibrating} by splitting the power spectrum into a {\it no-wiggle} ($P_{nw}$) and a {\it wiggle} ($P_\mathrm{lin}-P_{nw}$) component as
\begin{equation}\label{eq:pkmu}
    P(k,\mu) = (b+\mu^2 f)^2\left[ (P_\mathrm{lin}-P_\mathrm{nw})e^{-k^2\Sigma^2(\mu)}+P_\mathrm{nw}\right].
\end{equation}
The wiggle component is smoothed anisotropically through the damping scale, $\Sigma(\mu)$. In the previous expression, we also included the effect from the linear galaxy bias ($b$) and redshift space distortions ($\mu^2 f$) \cite{kaiser1987clustering}, with the latter proportional to the growth rate $f$. We model the no-wiggle component using a 1D Gaussian smoothing in log-space following Appendix A of \cite{vlah2016perturbation}.\footnote{\baofitwtheta also allows for using other implementations of the no-wiggle matter power spectrum, namely the ones available in \cosmoprimo (\url{https://github.com/cosmodesi/cosmoprimo.git}) (e.g., the one described in Appendix D of \cite{wallisch2019cosmological} or the one in Appendix D of \cite{brieden2022model}).} We also follow the infrared resummation model from \cite{senatore2015ir,blas2016time} to compute the damping scale $\Sigma(\mu)$ \cite{ivanov2020cosmological,ivanov2018infrared} relative to the line of sight (see details in \cite{y3-baokp}). This is consistent with the modeling used in the fiducial DES Y6 BAO analysis \cite{y6-baokp}.

Once we have the matter power spectrum, we decompose it into multipoles, $P_\ell(k)$, perform a Hankel transform to obtain the configuration space multipoles, $\xi_\ell(s)$, and then reconstruct the anisotropic redshift-space correlation function, $\xi(s,\mu)$ for each tomographic bin. From there, the angular correlation function of the $i$-th tomographic bin is obtained as
\begin{equation}\label{eq:w_template_schematic}
    w^i(\theta)=\int\dd{z_1}\int\dd{z_2}n^i(z_1)n^i(z_2)\xi^i\big( s(z_1,z_2,\theta),\mu(z_1,z_2,\theta)\big),
\end{equation}
where the $n^i(z)$ is the redshift distribution of that bin (we only calculate the galaxy clustering auto-correlations).

Our model contains the BAO template component $w_\mathrm{template}(\theta)$ described by \autoref{eq:w_template_schematic}, a BAO scaling parameter $\alpha$, an amplitude rescaling factor $A$ (per tomographic bin) and a smooth component $B(\theta)$ (also per tomographic bin):
\begin{equation}\label{eq:wtheta_model}
    w_\mathrm{model}^i(\theta,\boldsymbol{\Theta}) = A^i w_\mathrm{template}^i(\alpha\theta) + B^i(\theta). 
\end{equation}
Here, $\boldsymbol{\Theta}$ denotes the free parameters of our BAO model: $\alpha$, the amplitude $A$ and the broadband parameters $a_j$ included in $B(\theta)$, which we will discuss below. The parameter $\alpha$ quantifies the shift in the position of the BAO peak in the data with respect to the one in the template,
\begin{equation}
    \alpha = \frac{D_M(z)}{r_d}\frac{r_d^{\rm fid}}{D_M^{\rm fid}(z)},
\end{equation}
where $D_M(z)$ is the comoving angular diameter distance and $r_d$ is the sound horizon scale. The superscript ``fid'' indicates quantities evaluated at the fiducial cosmology of the template (for our analysis, we use two different cosmologies: \mice for the COLA mocks and \planck for the DES data,\footnote{We use \mice cosmology for the mocks so the template matches the underlying cosmological parameters of the simulations. For the data, we use \planck, since we expect the cosmology of the data to be closer to it. However, as shown in \cite{y6-baokp}, the choice of the cosmological parameters for the template does not bias our BAO distance measurements.} and their corresponding cosmological parameters are listed in \autoref{tab:cosmo_params}). The term $B(\theta)$ is introduced to absorb smooth components (not sharp features) that may come from remaining theoretical or observational systematic errors in the galaxy clustering signal. We model them as a sum of power laws,
\begin{equation}\label{eq:broadband_terms_w}
    B^i(\theta) = \sum_j\frac{a_j^i}{\theta^j},
\end{equation}
where $a_j^i$ are the broadband parameters of the $i$-th tomographic bin. We use $j = 0$, 1, 2 by default in this paper (see \cite{y6-baokp} for a more detailed discussion on this choice). The total number of free parameters ($\alpha$, $A^i$ and $a_j^i$) is, therefore,
\begin{equation}
    1 + 4 \times \text{(number of redshift bins)} = 25.
\end{equation}

Finally, the $\chi^2$ to be minimized in our BAO-fitting procedure is given by
\begin{equation}\label{eq:chi2_BAOfit}
    \chi^2(\boldsymbol{\Theta})=\sum_{i,j}\sum_{m,n}\Delta w^i(\theta_m,\boldsymbol{\Theta})(\text{cov}^{-1})_{m,n}^{i,j}\Delta w^j(\theta_n,\boldsymbol{\Theta}),
\end{equation}
where
\begin{equation}
    \Delta w^i(\theta,\boldsymbol{\Theta})=w_{\rm data}^i(\theta)-w_{\rm model}^i(\theta,\boldsymbol{\Theta}),
\end{equation}
with the model given by \autoref{eq:wtheta_model} and the covariance described in \autoref{sec:cov}. The indices $i,j$ denote tomographic bins, whereas the indices $m,n$ denote angular bins. Although the $\chi^2$ given by \autoref{eq:chi2_BAOfit} formally depends on both $\alpha$ and the nuisance parameters of the BAO model (amplitudes and broadband terms), in practice we follow a profile-likelihood procedure to obtain a $\chi^2$ that depends only on $\alpha$. For each value of $\alpha$, we determine the optimal amplitude parameters $A^i$ while analytically fitting the broadband parameters $a_j^i$. This is described in \cite{y6-baokp}.

\subsection{Covariance}\label{sec:cov}

The covariance is estimated using the \cosmolike code \cite{krause2017cosmolike,fang20202d,fang2020beyond}, which was also used for both the DES Y3 and Y6 BAO analyses \cite{y3-baokp,y6-baokp}. We stick to the use of Gaussian covariance matrices since we tested the impact of including non-Gaussian contributions, such as the trispectrum and the super-sample covariance terms, and found that they do not impact our results.

The covariance matrices require as inputs:
\begin{enumerate}
    \item Area of the angular mask (see \autoref{tab:areas_masks}).
    \item Number of galaxies per redshift bin (see \autoref{tab:ngal_samples}).
    \item Cosmological parameters. As mentioned earlier, for the COLA mocks, we use \mice cosmology (matching the underlying cosmology of the mocks). For the data, we use \planck. The cosmological parameters for both cosmologies are shown in \autoref{tab:cosmo_params}.
    \item Redshift distributions. For the COLA mocks, we use the average $n(z)$ of the 1,952 mocks, which reproduces very well the one of the data (see Fig. 2 of \cite{y6-baokp}). For the data, we use the fiducial $n(z)$ from the Y6 BAO analysis (shown in Fig. 4 of \cite{mena2024dark}).
    \item Linear galaxy bias per redshift bin. For the COLA mocks, we estimate them from the average of the 1,952 simulations. We perform this measurement only for the DES no-DESI and DES DESI COLA mocks, since the bias for the full-area mocks was already obtained in \cite{y6-baokp}. For the data, we measure them from the unblinded $w(\theta)$. In both cases, the measurement of the linear galaxy bias is performed in the theta range $0.5\,{\rm deg}<\theta<2\,{\rm deg}$.
\end{enumerate}

\subsection{BAO systematics}\label{sec:bao_sys}

We estimate the error budget when measuring the BAO signal for two different sources of systematics: 
\begin{itemize}
    \item Bias in $\alpha$ due to uncertainties in the redshift distributions. Since the redshift distributions enter the projection in redshift when computing the angular correlation function from the three-dimensional correlation function, uncertainties in them can slightly shift the inferred BAO angular position. In \cite{y6-baokp}, this effect was quantified by testing several redshift distributions, including two alternatives derived from our photo-$z$ algorithm, one from an overlapping spectroscopic sample, one from the clustering-$z$ technique, and the fiducial one, which effectively combines the three methods. This leads to an estimated systematic uncertainty of $\sigma_\mathrm{th,sys}^{\rm redshift} = 0.0045$ for the angular correlation function.\footnote{This value corresponds to the most extreme case, labeled ``All ACF'' in Table I of \cite{y6-baokp}.}
    \item Bias in $\alpha$ due to the non-linear evolution of the BAO peak. Non-linear evolution of the LSS predicts a shift in the BAO position of the order of $\delta\alpha\sim +0.5\%$ with respect to the linear case, with the exact value depending on the redshift range, linear bias and halo occupation distribution of the sample \cite{crocce2008nonlinear,padmanabhan2009calibrating}. In \cite{y6-baokp}, this effect was quantified from the COLA mocks, finding a systematic of $\sigma_\mathrm{th,sys}^{\rm modeling} = 0.0057$.\footnote{This value corresponds to the ``$i=0,1,2$'' entry in Table II of \cite{y6-baokp}.}
\end{itemize}
The total systematic budget from these two contributions is given by their sum in quadrature,
\begin{equation}\label{eq:bao_sys}
    \sigma_\mathrm{th,sys}^{\rm tot} = 0.0073.
\end{equation}
In \autoref{sec:bao_data}, we add this number in quadrature to the statistical error obtained from the BAO fits to compute the total error in $\alpha$.

\subsection{Differences with the Y6 BAO analysis}

In \autoref{tab:default_settings}, we show the default settings used in this paper for the BAO fits, which are the same as we used in the fiducial Y6 BAO analysis, see \cite{y6-baokp}. Besides the reduced area, the main difference with respect to the fiducial Y6 analysis is that here we only use the two-point angular correlation function, $w(\theta)$, as our galaxy clustering measurement.

While the fiducial DES Y6 analysis \cite{y6-baokp} combined three clustering statistics—the angular correlation function $w(\theta)$, the angular power spectrum $C_\ell$ \cite{camacho2019dark} and the projected correlation function $\xi_p(s)$ \cite{ross2017optimized,chan2022clustering,chan2022dark}—to maximize statistical power, in this work we focus exclusively on the angular correlation function. Calculating and validating the cross-covariance between different estimators (particularly Fourier-space statistics) becomes computationally expensive and methodologically complex when applied to the irregular geometry of the no-DESI footprint. By restricting to $w(\theta)$, we can conservatively apply the systematic error budget from \autoref{eq:bao_sys} rigorously characterized for the ACF in the Y6 analysis, ensuring robustness without the need to re-validate the full consensus framework on the reduced samples.\footnote{Based on the results in Table~VIII of \cite{y6-baokp}, this choice degrades the expected precision from 2.1\% to 2.4\%, corresponding to an approximately 14\% increase in the statistical uncertainty relative to the combined analysis.}

\begin{table*}[]
    \renewcommand{\arraystretch}{1.3}
    \setlength{\tabcolsep}{4pt}
    \centering
    \begin{tabular}{l|cc}
        \toprule\toprule
        & COLA mocks & DES BAO data \\
        \midrule
        \textbf{\boldmath Covariance \& Template $w(\theta)$} & & \\
        Cosmology & \mice & \planck \\
        Redshift distributions & Mean of mocks & Fiducial from \cite{y6-baokp} \\
        % Number of broadband parameters & $i = 0$, 1, 2 & $i = 0$, 1, 2 \\
        Galaxy bias & Measured from mean of mocks & Measured from data \\
        \midrule
        \textbf{Scales for the BAO fits} & & \\
        $\theta$ range (deg) & [0.5, 5] & [0.5, 5] \\
        $\theta$ binning (deg) & 0.2 & 0.2 \\
        \bottomrule\bottomrule
    \end{tabular}

    \caption{Default settings for the theoretical template and the covariance, and scales used for the BAO fits.}
    \label{tab:default_settings}
\end{table*}

\section{Measurement of the BAO}\label{sec:bao_measurements}

In this section, we present the BAO measurements for the no-DESI samples. We follow a blind procedure: first, in \autoref{sec:mocks_results}, we verify that the shifts in $\alpha$ and the increases in $\sigma_\alpha$ after removing the DESI overlap are consistent with mock statistics, considering only relative shifts and ratios rather than absolute values; second, in \autoref{sec:bao_data}, we present the unblinded measurements.

\subsection{Pre-unblinding tests}\label{sec:mocks_results}

Our goal here is to check whether the changes in $\alpha$ ($\Delta\alpha \equiv \alpha-\alpha_\mathrm{full\ sample}$) and the increases in $\sigma_\alpha$ ($R_{\sigma_\alpha} \equiv \sigma_\alpha/\sigma_\alpha^{\rm full\ sample}$) measured on the DES Y6 BAO sample when removing the area of overlap with DESI are consistent with the statistics of the COLA mocks (for which we do the same splits as in the data). Therefore, here we do not show the explicit values for $\alpha$ in the data or their errors, but rather differences or ratios with respect to the full sample.

In \autoref{tab:delta_alpha}, we display $\Delta\alpha$ for the different data splits, both for the BAO sample and the mocks. In the case of the mocks, we show the 90 and 99\% intervals defined by their statistics. We see that the shifts in $\alpha$ measured in the data are well within the 90\% regions defined by the mocks for all the different splits besides \decbelow, which has a relatively high shift (corresponding to the 97.7\% interval, still within the 99\% one). Both DES no-DESI samples show a preference for higher $\alpha$ relative to the full footprint, while the DES DESI samples prefer a lower value.
\begin{table}[]
    \renewcommand{\arraystretch}{1.3} % Default value: 1
    \setlength{\tabcolsep}{4pt} % Adjust column separation
    \centering
    \begin{tabular}{lccc}
        \toprule\toprule
        $(\alpha-\alpha_\mathrm{full\ sample})\times 100$ & Data & 90\%-mocks & 99\%-mocks \\
        \hline
        \tilesnodesi & 1.56 & [-1.88, 2.04] & [-3.50, 3.50] \\
        \tilesonlydesi & -4.29 & [-6.41, 6.24] & [-13.03, 9.67] \\\hline
        \decbelow & 2.92 & [-2.28, 2.28] & [-3.96, 3.82] \\
        \decabove & -2.73 & [-5.84, 5.52] & [-10.66, 9.47] \\
        \bottomrule\bottomrule
    \end{tabular}
    \caption{Difference in the best-fit $\alpha$ between the different splits and the full sample. We show the result on the data and the 90 and 99\% intervals from the COLA mocks.}
    \label{tab:delta_alpha}
\end{table}

In \autoref{tab:delta_sigma_alpha}, we display $R_{\sigma_\alpha}$ for the different data splits, both for the data and for the mocks. We find that the results on the data are consistent with the 90\% intervals defined by the mocks for all the cases. The increases in $\sigma_\alpha$ obtained on the data when removing regions of the footprint are consistent with the ones we would expect from the re-scaling of the area (i.e., computing $A_\mathrm{full}/A$ from the data displayed in \autoref{tab:areas_masks}).
\begin{table}[]
    \renewcommand{\arraystretch}{1.3} % Default value: 1
    \setlength{\tabcolsep}{4pt} % Adjust column separation
    \centering
    \begin{tabular}{lcccZ}
        \toprule\toprule
        $\sigma_\alpha/\sigma_\alpha^{\rm full\ sample}$ & Data & 90\%-mocks & 99\%-mocks & Expected \\
        \hline
        \tilesnodesi & 1.32 & [1.02, 1.41] & [0.94, 1.69] & 1.15 \\
        \tilesonlydesi & 2.86 & [1.76, 3.86] & [1.41, 5.74] & 2.02 \\\hline
        \decbelow & 1.40 & [1.06, 1.55] & [0.94, 1.97] & 1.21 \\
        \decabove & 2.11 & [1.53, 3.19] & [1.24, 4.25] & 1.79 \\
        \bottomrule\bottomrule
    \end{tabular}
    \caption{Ratio of $\sigma_\alpha$ between the different sample splits and the full sample. We show the result on the data and the 90 and 99\% intervals from the COLA mocks.}
    \label{tab:delta_sigma_alpha}
\end{table}

All these tests demonstrate that the observed shifts in $\alpha$ and $\sigma_\alpha$ are consistent with statistical fluctuations expected from the COLA mocks, with no indication of systematic biases induced by removing the DESI-overlap region. Based on these results, we proceeded to the unblinding of the BAO measurements.

\subsection{Unblinded measurements}\label{sec:bao_data}

In \autoref{tab:robustness}, we present the unblinded BAO measurements for the various data splits. The table includes measurements obtained using the fiducial settings (``Fiducial'' column, which applies the LSS systematics weights and uses $\theta_{\rm min}=0.5$ deg and $\Delta\theta=0.2$ deg), as well as those derived without applying the LSS weights (``no-$w_\mathrm{sys}$'' column), and with modified scale cuts (``$\theta_{\rm min}=1$ deg'', ``$\Delta\theta=0.05$ deg'' and ``$\Delta\theta=0.1$ deg'' columns). The quoted uncertainties already include the contribution from BAO systematics as defined in \autoref{eq:bao_sys}. We find that the best-fit $\alpha$ values and their uncertainties remain stable when the imaging systematics weights are omitted—which primarily affect the amplitude of the two-point correlation function but not the BAO position—and when the scale cuts are varied, showcasing the robustness of our measurements.

We note that the fiducial result for the full sample ($0.9534 \pm 0.0228$) is consistent to within $<0.1\sigma$ with the fiducial Y6 BAO analysis of the ACF ($0.9517 \pm 0.0227$, from Table VIII of \cite{y6-baokp}), as expected given that we employ the same analysis pipeline with only minor updates. 

The uncertainty in $\alpha$ obtained for each sample follows the expected trend: \tilesnodesi yields a slightly smaller error than \decbelow, consistent with its larger area (see \autoref{tab:areas_masks}). Conversely, \tilesonlydesi shows a slightly larger error than \decabove, reflecting its smaller area.

The BAO measurement adopted in \autoref{sec:cosmo_constraints} for the cosmological parameter inference corresponds to the \tilesnodesi sample ($\alpha = 0.9690\pm0.0296$, highlighted in bold). As we already discussed in the previous section, the best-fit value of $\alpha$ for \tilesnodesi increases by 1.56\% relative to the full sample. Since the uncertainty in $\alpha$ is larger due to the smaller area, this measurement is consistent within $\sim$$1\sigma$ with the prediction from \planck ($\alpha = 1$), whereas the full sample shows a $\sim$$2\sigma$ tension. Conversely, the $\alpha$ value for the \tilesonlydesi sample, which is complementary to \tilesnodesi, decreases with respect to the full sample, such that their weighted average remains consistent with the fiducial result.

\begin{table*}[]
    \renewcommand{\arraystretch}{1.3}
    \setlength{\tabcolsep}{6pt}
    \centering
    \begin{tabular}{l|c|c|c|c|c}
        \toprule\toprule
        Dataset &
        Fiducial &
        no-$w_\mathrm{sys}$ &
        $\theta_\mathrm{min}=1$ deg &
        $\Delta\theta=0.05$ deg &
        $\Delta\theta=0.1$ deg \\
        \midrule
        Full sample &
        $0.9534 \pm 0.0228$ &
        $0.9558 \pm 0.0232$ &
        $0.9514 \pm 0.0228$ &
        $0.9497 \pm 0.0228$ &
        $0.9506 \pm 0.0228$ \\\hline

        \tilesnodesi &
        \boldmath $0.9690 \pm 0.0296$ &
        $0.9742 \pm 0.0304$ &
        $0.9618 \pm 0.0287$ &
        $0.9646 \pm 0.0274$ &
        $0.9662 \pm 0.0280$ \\

        \tilesonlydesi &
        $0.9105 \pm 0.0627$ &
        $0.9177 \pm 0.0625$ &
        $0.9041 \pm 0.0636$ &
        $0.9197 \pm 0.0646$ &
        $0.9197 \pm 0.0641$ \\\hline

        \decbelow &
        $0.9826 \pm 0.0313$ &
        $0.9890 \pm 0.0313$ &
        $0.9786 \pm 0.0317$ &
        $0.9798 \pm 0.0301$ &
        $0.9810 \pm 0.0306$ \\

        \decabove &
        $0.9261 \pm 0.0465$ &
        $0.9301 \pm 0.0461$ &
        $0.9201 \pm 0.0488$ &
        $0.9325 \pm 0.0476$ &
        $0.9329 \pm 0.0473$ \\
        \bottomrule\bottomrule
    \end{tabular}
    \caption{Results of the BAO fits ($\alpha$) on the DES BAO data for the different splits. For each case, we display the fiducial result (which applies the LSS systematics weights and uses $\theta_{\rm min}=0.5$ deg and $\Delta\theta=0.2$ deg) and the ones obtained when not applying the LSS weights and when changing the scale cuts. The errors already contain the contribution from systematics given by \autoref{eq:bao_sys}. The highlighted case (\tilesnodesi) is the DES BAO likelihood we use for the cosmological analysis in \autoref{sec:cosmo_constraints}. The values of $\alpha$ and their uncertainties are reported to four decimal places, matching the convention of \cite{y6-baokp} and allowing small variations between different cases to be more easily discerned. The fiducial result for the full sample ($0.9534 \pm 0.0228$) is consistent to within $<0.1\sigma$ with the fiducial Y6 BAO result for the ACF ($0.9517 \pm 0.0227$, from Table VIII of \cite{y6-baokp}). }
    \label{tab:robustness}
\end{table*}

\begin{figure}
    \centering
    \includegraphics[width=\linewidth]{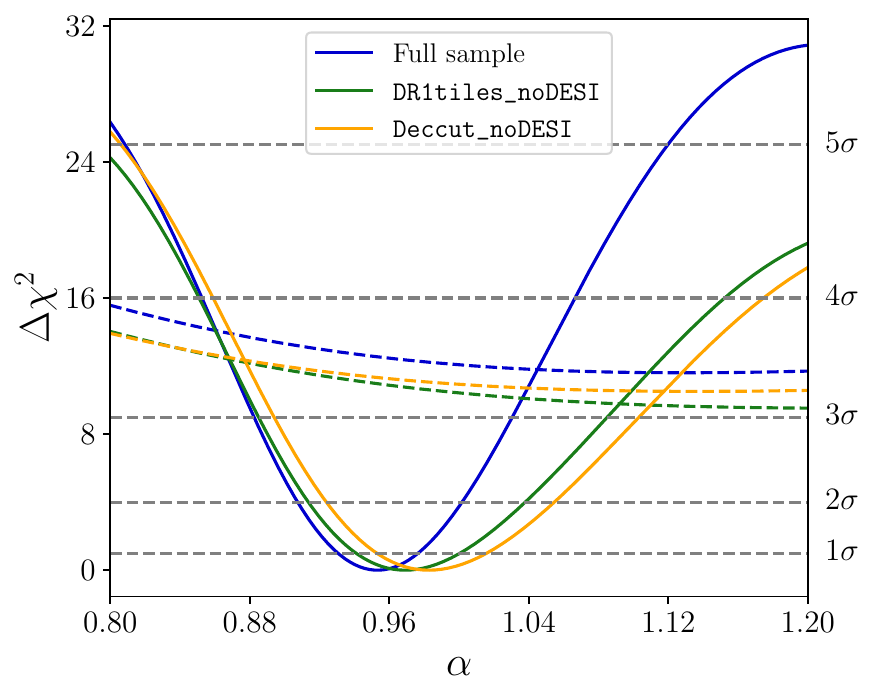}
    \caption{$\Delta\chi^2$ profile for the full sample and the two no-DESI samples (\tilesnodesi and \decbelow), given by \autoref{eq:chi2_BAOfit} following the minimization procedure described in \cite{y6-baokp}. The three curves have been stretched to account for the systematic uncertainty given by \autoref{eq:bao_sys}. In colored dashed lines, we show the $\Delta\chi^2$ obtained when trying to fit the data with a template without BAO. The 1, 2, 3, 4 and 5$\sigma$ limits are shown as horizontal black dashed black lines. We see that all the cases have a significance for the detection of the BAO peak within 3 and 4$\sigma$, being the full sample case the highest one, as expected.}
    \label{fig:likelihoods_all}
\end{figure}

In \autoref{fig:likelihoods_all}, we show the $\Delta\chi^2(\alpha)\equiv\chi^2(\alpha)-\chi^2_{\rm min}$ as a function of $\alpha$ for \tilesnodesi and \decbelow. We also include the full sample for comparison. The three curves have been stretched to account for the systematic uncertainty given by \autoref{eq:bao_sys}. The colored dashed lines show the $\Delta\chi^2$ obtained when fitting the data with a template without BAO, and we use them to estimate the significance of the detection of the BAO peak. We see that for both no-DESI samples, the significance is a bit lower than for the full sample case (as expected), but it is still between 3 and 4$\sigma$ in the entire $\alpha$ range.

\begin{figure}
    \centering
    \includegraphics[width=\linewidth]{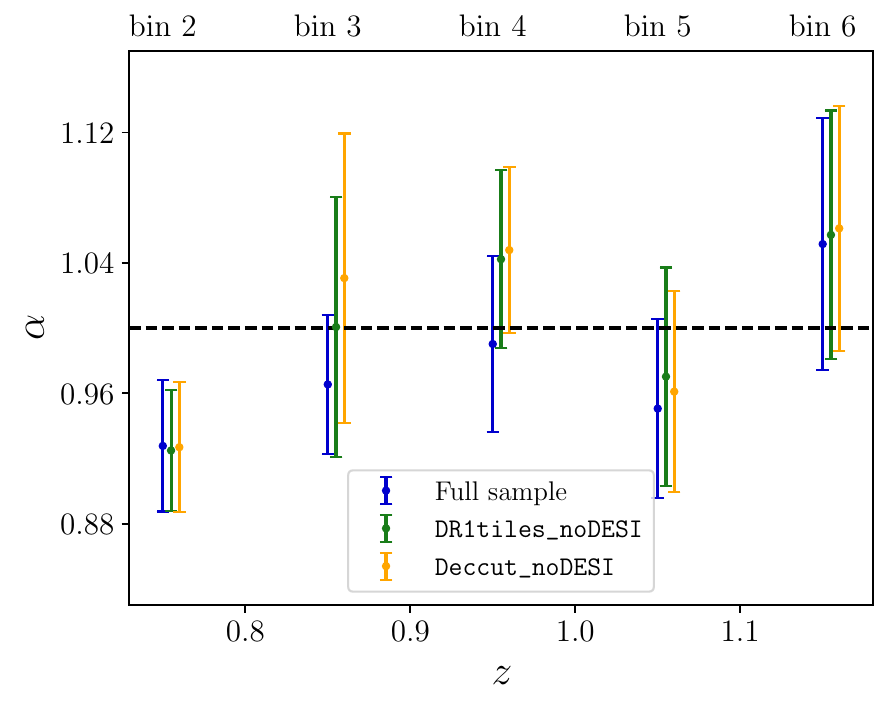}
    \caption{Results of the BAO fits for the individual redshift bins for the full sample and the two no-DESI samples (\tilesnodesi and \decbelow). Based on our BAO-detection criteria (see \cite{y6-baokp}), bin 1 has a non-detection and, therefore, it is omitted from the plot.}
    \label{fig:bao_individual_bins}
\end{figure}

In \autoref{fig:bao_individual_bins}, we show the results of the BAO fits for the individual redshift bins for the two no-DESI samples (\tilesnodesi and \decbelow), and also the full sample for comparison. The first redshift bin (labeled as ``bin 1'' but not included in the figure) has a non-detection of the BAO for both no-DESI samples, similarly to the full sample (which is well explained by the statistics of the mocks, see \cite{y6-baokp}), and therefore it is omitted from the plot. We find that the measurements in bins 2, 5 and 6 are consistent between the three samples. However, the full sample has the lowest $\alpha$ in bins 3 and 4 among the three, which explains why its best-fit alpha when fitting all bins together is the lowest (see \autoref{fig:likelihoods_all}). On the other hand, \decbelow has the largest $\alpha$ in these two bins, which is consistent with it having the largest $\alpha$ when fitting all bins together.

\section{Cosmological constraints}\label{sec:cosmo_constraints}

In this section, we describe the methodology used for the cosmological inference, summarize the datasets employed, and present the resulting cosmological constraints and associated levels of tension with \lcdm when interpreting the data within the \wacdm framework.

\subsection{Methods}

Here, we outline the procedures used for the cosmological inference and for quantifying the tension with \lcdm.

\subsubsection{Parameter inference}\label{sec:inference}

To derive constraints on the parameters $\mathbf{p}$ given the data $\mathbf{D}$ under the model $\mathcal{M}$, we construct the posterior probability distribution, $\mathcal{P}(\mathbf{p}\, |\, \mathbf{D}, \mathcal{M})$, according to Bayes' theorem as
\begin{equation}\label{eq:posterior}
    \mathcal{P}(\mathbf{p}\, |\, \mathbf{D}, \mathcal{M}) = \frac{\mathcal{L}(\mathbf{D}\,|\, \mathbf{p}, \mathcal{M} ) \, \Pi(\mathbf{p}\,|\,\mathcal{M})}{\mathcal{Z}(\mathbf{D}\,|\,\mathcal{M})},
\end{equation}
where $\Pi(\mathbf{p}\,|\,\mathcal{M})$ is the prior probability distribution for the parameters, $\mathcal{L}(\mathbf{D}\,|\, \mathbf{p}, \mathcal{M} )$ is the likelihood of observing the data given the model parameters and $\mathcal{Z}(\mathbf{D}\,|\,\mathcal{M})$ is the Bayesian evidence, defined as
\begin{equation}\label{eq:evidence}
    \mathcal{Z}(\mathbf{D}\,|\,\mathcal{M}) = \int \mathrm{d}\mathbf{p} \, \mathcal{L}(\mathbf{D}\,|\, \mathbf{p}, \mathcal{M} ) \, \Pi(\mathbf{p}\,|\,\mathcal{M}).
\end{equation}
Assuming Gaussianity, the likelihood can be written as
\begin{equation}
    \mathcal{L}(\mathbf{D}\,|\,\mathbf{p}, \mathcal{M}) \propto 
    \exp\!\left[-\frac{1}{2}\,\chi^2(\mathbf{D}\,|\,\mathbf{p}, \mathcal{M})\right],
\end{equation}
where the $\chi^2$ function quantifies the goodness of fit between the model $\mathcal{M}$ and the data $\mathbf{D}$:
\begin{equation}
    \chi^2(\mathbf{D}\,|\,\mathbf{p}, \mathcal{M}) =
    \big[\mathbf{D} - \mathbf{t}(\mathbf{p}, \mathcal{M})\big]^\mathrm{T}
    \mathbf{C}^{-1}
    \big[\mathbf{D} - \mathbf{t}(\mathbf{p}, \mathcal{M})\big].
\end{equation}
Here, $\mathbf{C}$ denotes the covariance matrix of the measurements, and $\mathbf{t}(\mathbf{p}, \mathcal{M})$ is the theoretical prediction for the observables given the cosmological parameters $\mathbf{p}$ and the model $\mathcal{M}$. Throughout this work, we utilize the likelihoods and $\chi^2$ functions for each of the datasets described in \autoref{sec:data}. We treat the likelihoods of different datasets as uncorrelated.

The complete list of sampled parameters and their prior ranges is provided in \autoref{tab:priors}. Additionally, when working with the \wacdm model, we impose the prior condition $w_0 + w_a < 0$, ensuring that $w(a) < 0$ at all redshifts and thereby excluding parameter regions incompatible with a radiation-dominated early universe.

We sample the posterior distributions using the \nautilus sampler \cite{lange2023nautilus}, a modern nested sampling algorithm designed for efficient exploration of high-dimensional and potentially multi-modal posterior distributions. Our analysis is implemented within the \cosmosis\footnote{\url{https://cosmosis.readthedocs.io/}.} framework \cite{zuntz2015cosmosis}, which serves as our primary inference pipeline. We compute the theoretical predictions with the \camb Boltzmann solver \cite{howlett2012cmb,lewis2000efficient} and use the \halofit Takahashi model \cite{takahashi2012revising,bird2012massive} for the nonlinear matter power spectrum when analyzing the CMB likelihood.

Parameter constraints are reported as the mean values, and we use the \getdist\footnote{\url{https://github.com/cmbant/getdist}.} package \cite{Lewis_2025} to compute equal-posterior credible regions and visualize the posterior distributions. We examine the 68\% credible regions and use their distance from the mean to determine (potentially asymmetric) $1\sigma$ errors, following the exact same methodology as in \cite{abbott2025dark}.

\begin{table}
    \renewcommand{\arraystretch}{1.2} % Default value: 1
    \setlength{\tabcolsep}{15pt} % Adjust column separation
    \centering	
    \begin{tabular}{cZc}
        \toprule\toprule
        Parameter & Fiducial & Prior \\\hline
    
        \multicolumn{3}{c}{\bf \boldmath\lcdm} \\
        $\hubble$ [\kmsMpc] & 69 & [55, 91] \\
        $\om$ & 0.3 & [0.1, 0.9] \\ 
        $\ob$ & 0.048 & [0.03, 0.07] \\
        $\tau$ & 0.067 & [0.04, 0.15]\\
        $A_\mathrm{s} \times 10^{9}$ & 2.19 & [0.5, 5.0] \\ 
        $n_\mathrm{s}$ & 0.97 & [0.87, 1.07] \\
        $a_\mathrm{Planck}$ & 1.0 & (1.0, 0.0025) \\
        \hline
    
        \multicolumn{3}{c}{\bf \boldmath\wacdm} \\
        $w_0$ & -1.0 & [-3, -0.33] \\
        $w_a$ & 0.0 & [-3, 3] \\

        \bottomrule\bottomrule
    \end{tabular}
    \caption{Sampled parameters and priors used in the  \lcdm and \wacdm analyses. The parameters $\tau$, $A_\mathrm{s} \times 10^{9}$, $n_\mathrm{s}$ and $a_\mathrm{Planck}$ are varied in all the chains used for this analysis since all of them include CMB from Planck. Square brackets denote a flat prior, while parentheses denote a Gaussian prior of the form $\mathcal{N}(\mu,\sigma)$, with $\mu$ and $\sigma$ being the mean and standard deviation, respectively. The parameter \neutrinomass is fixed to 0.06 eV.} 
    \label{tab:priors}
\end{table}

\subsubsection{Quantifying deviations from \lcdm}\label{sec:quantifying}

To estimate the preference for the \wacdm model relative to \lcdm, we compute the difference in the best-fit $\chi^2$ values between the two models (for a given dataset or combination of datasets).\footnote{Unlike in \cite{abbott2025dark}, in the present analysis we do not use \tensiometer (\url{https://github.com/mraveri/tensiometer}) to evaluate the tension with \lcdm when interpreting our datasets under \wacdm, but differences in best-fit $\chi^2$ instead. However, we verified that both methods give consistent results.} For the chains that include the CMB likelihood, there are several additional parameters compared to those that do not include it (see \autoref{tab:priors}). We note that $a_\mathrm{Planck}$ has a Gaussian prior, which makes it difficult to directly maximize the likelihood (which does not encode information about the priors), as the posterior might prefer a value for this parameter that is too far from the Gaussian prior and yields a higher likelihood. For this reason, we maximize the posterior instead, which from \autoref{eq:posterior} can be written as
\begin{equation}\label{eq:posterior_priorsplit}
    \mathcal{P}(\mathbf{p}\, |\, \mathbf{D}, \mathcal{M}) = 
    \frac{\mathcal{L}(\mathbf{D}\,|\,\mathbf{p},\mathcal{M})\, 
    \Pi_\mathrm{G}(\mathbf{p}\,|\,\mathcal{M})\,
    \Pi_\mathrm{flat}(\mathcal{M})}{\mathcal{Z}(\mathbf{D}\,|\,\mathcal{M})},
\end{equation}
where $\Pi_\mathrm{G}(\mathbf{p}\,|\,\mathcal{M})$ denotes the Gaussian priors and $\Pi_\mathrm{flat}(\mathcal{M})$ the flat priors. Since $\Pi_\mathrm{flat}(\mathcal{M})$ does not depend on the parameters $\mathbf{p}$, the best fit for $\mathcal{L}(\mathbf{D}\,|\, \mathbf{p}, \mathcal{M} ) \, \Pi_\mathrm{G}(\mathbf{p}\,|\,\mathcal{M})$ will be the same vector of parameters $\mathbf{p}_\mathrm{MAP}$ that maximizes the posterior. The vector of parameters $\mathbf{p}_\mathrm{MAP}$ is searched in the following way: we identify the $50$ samples with the highest posterior in the original chain file, and use these as starting points for a local maximization of the posterior employing the Powell algorithm \citep{powell1964efficient}, a derivative-free optimization method.

Evaluating \autoref{eq:posterior_priorsplit} at $\mathbf{p}=\mathbf{p}_\mathrm{MAP}$, taking logarithms on both sides, and defining
\begin{align}
        \left[\chi^2_{P}\right]_\mathcal{M} & \equiv -2 \log \mathcal{P}(\mathbf{p}_\mathrm{MAP}\, |\, \mathbf{D}, \mathcal{M}), \\
        \left[\chi^2_{\mathcal{L}\Pi_\mathrm{G}}\right]_\mathcal{M} & \equiv -2 \log \left(\mathcal{L}(\mathbf{D}\,|\,\mathbf{p}_\mathrm{MAP},\mathcal{M})\Pi_\mathrm{G}(\mathbf{p}_\mathrm{MAP}\,|\,\mathcal{M})\right),
\end{align}
we get
\begin{equation}
    \left[\chi^2_\mathcal{P}\right]_\mathcal{M} = \left[\chi^2_{\mathcal{L}\Pi_\mathrm{G}}\right]_\mathcal{M} - 2 \log \Pi_\mathrm{flat}(\mathcal{M}),
\end{equation}
which represents a $\chi^2$-like posterior-based goodness-of-fit for model $\mathcal{M}$. The difference in $\chi^2$ we use for the model comparison in this paper is, then, given by
\begin{multline}
    \Delta\chi^2 = \left[\chi^2_\mathcal{P}\right]_{\wacdm} - \left[\chi^2_\mathcal{P}\right]_{\lcdm} \\
    + 2 \log \Pi_\mathrm{flat}(\wacdm) - 2 \log \Pi_\mathrm{flat}(\lcdm).
\end{multline}
Taking the priors from \autoref{tab:priors}, we get
\begin{equation}
    \Delta\chi^2 = \left[\chi^2_\mathcal{P}\right]_{\wacdm} - \left[\chi^2_\mathcal{P}\right]_{\lcdm} - 2 \log(16.02).
\end{equation}
This correction removes the Bayesian prior volume factors to recover the frequentist profile likelihood ratio, allowing direct comparison with DESI methodology.

Finally, to translate $\Delta\chi^2$ into the statistical significance of this improvement, we quote the corresponding frequentist significance $N_\sigma$ for a 1D Gaussian distribution,
\begin{equation}\label{eq:nsigma}
    \mathrm{CDF}_{\chi^2}\left(\Delta\chi^2 |\, 2\,\mathrm{dof}\right) = \frac{1}{\sqrt{2\pi}}\int_{-N_\sigma}^{N_\sigma} e^{-t^2/2} \dd{t}, 
\end{equation}
where the left-hand side denotes the cumulative distribution of the $\chi^2$ function. This is equivalent to what was done by the DESI collaboration in \cite{desidr2bao}.

% We then compute the best-fit $\chi^2$ from the posterior as
% \begin{equation}
%     \chi^2_\mathrm{MAP}(\mathbf{D},\mathcal{M}) =
%     -2 \log P(\mathbf{p}_\mathrm{MAP}\,|\,\mathbf{D},\mathcal{M})
%     + 2 \log \pi_\mathrm{flat}(\mathcal{M}),
% \end{equation}
% where the last term removes the constant contribution from the flat priors, ensuring that $\chi^2_\mathrm{MAP}$ corresponds to the likelihood augmented only by Gaussian priors.

\subsection{Datasets}\label{sec:data}

Here, we describe the different datasets used for the cosmological inference in this analysis.

\subsubsection{DES BAO}

Our main DES BAO likelihood for this analysis comes from the \tilesnodesi sample, and is shown in green in \autoref{fig:likelihoods_all}. The best-fit $\alpha$ can be found in \autoref{tab:robustness} (fiducial case),
\begin{equation}
    \alpha(0.851) = 0.9690 \pm 0.0296,
\end{equation}
which already contains the contribution from statistical and systematic errors. Equivalently,
\begin{equation}
    D_M(0.851)/r_d = 19.74 \pm 0.60.
\end{equation}
Hereafter, we refer to it as ``DES BAO-noDESI'', or ``BAO-noDESI'' for simplicity. It is important to note that analyses based on DES BAO data alone (i.e., without combining with DESI) should use the fiducial Y6 likelihood, as it yields a more precise angular BAO measurement. This one is displayed in Eq. 34 of \cite{y6-baokp}: $\alpha(0.851) = 0.9571 \pm 0.0201$. In this paper, we also use this BAO likelihood for comparison purposes, and we refer to it as ``BAO-full''.

\subsubsection{DES SN}\label{sec:SN}

Since this paper is a follow-up to \cite{abbott2025dark}, we adopt the DES Y5 SN sample \cite{des-y5-sn} as our baseline supernova dataset. However, in the next section, we also describe the newly calibrated DES SN–Dovekie sample \cite{popovic2025dark}, which we likewise use in our cosmological inference.

DES released in 2024 the Hubble diagram based on Type Ia supernovae discovered over five years by the DES-SN program \cite{des-y5-sn}. Over 30,000 supernova candidates were identified using photometric data, from which 1,635 high-quality SN Ia-like events with spectroscopic host-galaxy redshifts were selected. These are included in the Hubble diagram with weights reflecting their Type Ia classification probabilities as estimated by \texttt{SuperNNova} \cite{moller2020supernnova,moller2022dark,vincenzi2024dark}. Additionally, 194 low-redshift ($z < 0.1$) spectroscopically confirmed Type Ia SN from external surveys are incorporated. Both samples are combined and referred to as ``SN'' in the cosmological analysis. All SN distances and their associated covariance matrices are publicly available.\footnote{\url{https://github.com/des-science/DES-SN5YR}.}

The DES Y5 SN analysis builds upon the DES Y3 SN study \cite{abbott2019first}, with details on photometric calibration, light-curve fitting and standardization given in \cite{brout2022pantheon+,sanchez2024dark,taylor2023salt2,vincenzi2024dark}. Host galaxy properties are described in \cite{kelsey2023concerning}, and systematic uncertainties are detailed in \cite{vincenzi2024dark}.

The standardized SN distance modulus is given by
\begin{equation}
    \mu_\mathrm{obs} = m_x + \alpha x_1 - \beta c + \gamma G_\mathrm{host}(M_\star) - M - \Delta\mu_\mathrm{bias},
    \label{eq:SN}
\end{equation}
where $m_x$, $x_1$, and $c$ are light-curve parameters, $\alpha$, $\beta$, and $\gamma$ are global nuisance parameters, and $G_\mathrm{host}(M_\star)$ accounts for host galaxy mass effects. The absolute magnitude $M$ is degenerate with the Hubble constant $H_0$ and is combined into a single parameter $\mathcal{M} = M + 5\log_{10}{(c/\hubble)}$ that is analytically marginalized over in the SN likelihood \cite{des-y5-sn}.

\subsubsection{DES SN-Dovekie}

The DES Supernova Program has recently produced an updated Type Ia Hubble diagram \cite{popovic2025dark}, hereafter referred to as the DES SN-Dovekie sample. This re-analysis of the five-year DES-SN program includes an improved photometric cross-calibration (incorporating white-dwarf standards), retraining of the SALT3 light-curve model and correction of a numerical approximation in the host-galaxy colour‐law. The resulting sample comprises 1,623 likely SN Ia from DES (compared to the 1,635 from \cite{des-y5-sn}) and 197 low-redshift spectroscopically confirmed SN from external surveys (compared to the 194 from \cite{des-y5-sn}). This updated dataset uses the same formalism as previous DES SN releases but benefits from the enhanced calibration and improved systematics modeling.

\subsubsection{Planck CMB}\label{sec:CMB}

We use the same Planck CMB likelihood as in our previous analysis \cite{abbott2025dark}. We incorporate measurements of the CMB temperature and polarization anisotropies using the Planck 2018 likelihood \cite{aghanim2020planckV}, which we refer to simply as ``CMB'' throughout this work. For multipoles $\ell \geq 30$, we use the {\tt Plik-lite} likelihood, which marginalizes over Planck foregrounds and nuisance parameters and includes data up to $\ell_\mathrm{max}=2,508$ for TT, and up to $\ell_\mathrm{max}=1,996$ for TE and EE spectra. At low multipoles ($2 \leq \ell < 30$), we follow the standard Planck approach, employing the {\tt Commander} likelihood for the TT spectrum and {\tt SimAll} for EE polarization. CMB lensing constraints are not included in our analysis.

Including the CMB likelihood requires us to extend the parameter space beyond that used in background-only studies. In particular, we fit the amplitude $A_\mathrm{s}$ and spectral index $n_\mathrm{s}$ of the primordial curvature power spectrum, as well as the optical depth to reionization, $\tau$. We also marginalize over a global Planck calibration parameter, $a_\mathrm{Planck}$, treated as a nuisance parameter. The priors for these quantities are listed in \autoref{tab:priors}.

\subsubsection{DESI BAO}

As this paper is a follow-up to \cite{abbott2025dark}, we include BAO measurements from both DESI DR1 \cite{desi2024iii} and the newer DESI DR2 \cite{desidr2bao}. Although DR1 has been superseded by DR2, it was the dataset used in \cite{abbott2025dark}, where we found that ignoring correlations and adding DES BAO-full to the combination SN+CMB+DESI-DR1 increased the preference for \wacdm from $3.7\sigma$ to $4.0\sigma$. Including DR1 here allows us to revisit that analysis and assess how this preference changes when replacing DES BAO-full with DES BAO-noDESI.

The DESI DR1 and DR2 BAO analyses provide robust measurements of the BAO distance scale from the anisotropic clustering of galaxies in configuration space, using the first year and the first three years of DESI data, respectively. The BAO feature is extracted from the two-point correlation functions of different galaxy tracers, yielding measurements of the comoving angular diameter distance relative to the sound horizon scale, $D_M(z)/r_d$, and the product of the Hubble parameter and the sound horizon scale, $H(z)r_d$, at multiple effective redshifts.

\begin{figure}
    \centering
    \includegraphics[width=\linewidth]{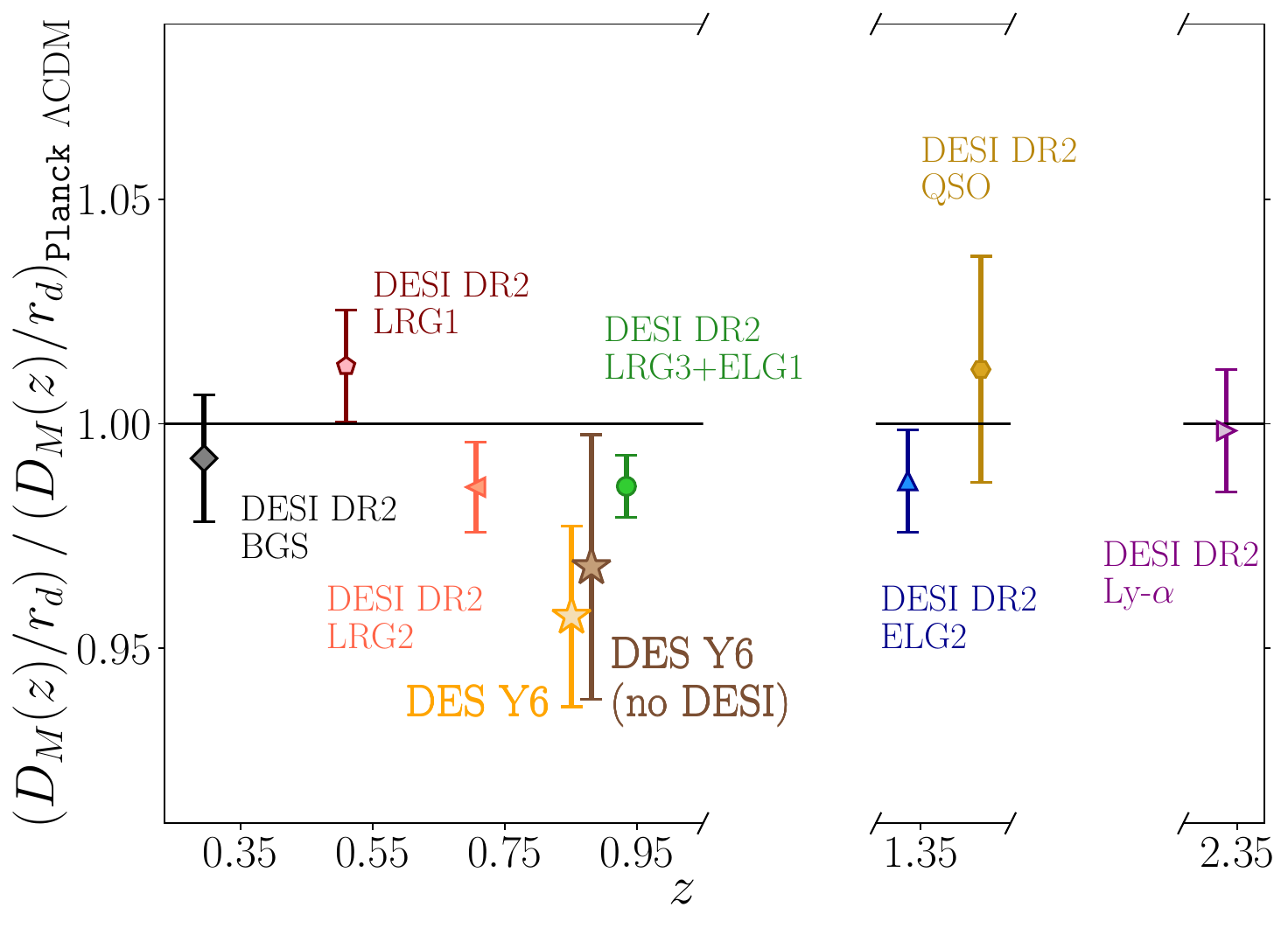}
    \caption{Ratio between the $D_M(z)/r_d$ measured using the BAO feature at different redshifts and the prediction from the cosmological parameters determined by Planck-2018, assuming \lcdm (black line). We include all the measurements from the DESI DR2 BAO analysis in different colors, the fiducial DES Y6 BAO measurement as a golden star and the DES no-DESI BAO measurement as a brown star (from the \tilesnodesi sample). We shifted the redshift of the latter so that it does not overlap with the measurement on the full sample, but they share the same effective redshift.}
    \label{fig:bao_distance_ladder}
\end{figure}

In \autoref{fig:bao_distance_ladder}, we show the angular BAO distance ladder, including the results from DESI DR2, our BAO-full measurement and the BAO-noDESI one (\tilesnodesi sample). In the DESI DR2 analysis, the Bright Galaxy Survey (BGS) tracer was analyzed using a 1D BAO fit, whereas all other tracers used a 2D fit. Therefore, for BGS we present $D_V/r_d$ in place of $D_M/r_d$ in the plot, and its uncertainty is shown as $1.5 \times \sigma(D_V/r_d)$.\footnote{Assuming spherical symmetry, the $D_M(z)$ constraints are 50\% less precise with respect to the spherically-averaged measurement, see \cite{ross2015information}.} However, for the chains we use the full DESI DR2 BAO likelihood, considering both $D_M/r_d$ and $D_H/r_d$ when available, and $D_V/r_d$ for BGS. As we discussed earlier, the no-DESI measurement has a higher $\alpha$ and a larger uncertainty (due to the reduced area) compared to the full sample, making it consistent within $\sim$$1\sigma$ with the prediction using \planck cosmology (black horizontal line).

\subsection{Results}

Here, we present the cosmological constraints under \lcdm and \wacdm, with particular emphasis on how they change when incorporating our new DES BAO-noDESI likelihood. We also quantify the level of preference for \wacdm relative to the standard \lcdm model. The results are organized into three subsections: (1) \autoref{sec:combination_dr1}, which uses DESI DR1 (follow-up of \cite{abbott2025dark}); (2) \autoref{sec:combination_dr2}, which updates to DESI DR2; and (3) \autoref{sec:dovekie}, which employs the DES SN–Dovekie dataset.

\subsubsection{Combination with DESI DR1}\label{sec:combination_dr1}

In \autoref{tab:constraints_cosmo_params_dr1}, we display the constraints on cosmological parameters, in both \lcdm and \wacdm, for various combinations of datasets that include DESI DR1.

\begin{table*}[]
    \centering
    \renewcommand{\arraystretch}{1.3}
    \setlength{\tabcolsep}{3.5pt}
    \begin{tabular}{l |c|c|c|c|c}
        \toprule\toprule
        Dataset & $\hubble$ (\kmsMpc) & \om & $\ob$ & $w_0$ & $w_a$ \\
        \midrule
        \multicolumn{6}{c}{\lcdm} \\
        \midrule
        % DESI-DR1 & -- & $0.2947^{+0.0136}_{-0.0152}$ & $> 0.0324$ & -- & -- \\
        % BAO-noDESI + DESI-DR1 & -- & $0.2936^{+0.0135}_{-0.0151}$ & $> 0.0324$ & -- & -- \\
        SN + CMB + DESI-DR1 & $67.81^{+0.40}_{-0.40}$ & $0.3091^{+0.0054}_{-0.0053}$ & $0.0488^{+0.0005}_{-0.0005}$ & -- & -- \\
        BAO-noDESI + SN + CMB + DESI-DR1 & $67.86^{+0.39}_{-0.38}$ & $0.3084^{+0.0052}_{-0.0052}$ & $0.0488^{+0.0004}_{-0.0004}$ & -- & -- \\
        \midrule
        \multicolumn{6}{c}{\wacdm} \\
        \midrule
        % DESI-DR1 & -- & $0.3344^{+0.0375}_{-0.0207}$ & $> 0.0300$ & $> -1.110$ & $< 0.498$ \\
        % BAO-noDESI + DESI-DR1 & -- & $0.3335^{+0.0379}_{-0.0200}$ & $> 0.0300$ & $> -1.127$ & $< 0.509$ \\
        SN + CMB + DESI-DR1 & $67.24^{+0.66}_{-0.65}$ & $0.3164^{+0.0066}_{-0.0067}$ & $0.0495^{+0.0010}_{-0.0010}$ & $-0.725^{+0.067}_{-0.074}$ & $-1.06^{+0.36}_{-0.29}$ \\
        BAO-noDESI + SN + CMB + DESI-DR1 & $67.28^{+0.66}_{-0.65}$ & $0.3160^{+0.0067}_{-0.0067}$ & $0.0495^{+0.0010}_{-0.0010}$ & $-0.720^{+0.068}_{-0.068}$ & $-1.00^{+0.35}_{-0.29}$ \\
        \bottomrule\bottomrule
    \end{tabular}
    % \caption{68\% credible region (1$\sigma$) or 95\% upper/lower limits of cosmological parameters under \lcdm and \wacdm for different data combinations including DESI DR1.}
    \caption{68\% credible region (1$\sigma$) of cosmological parameters under \lcdm and \wacdm for different data combinations including DESI DR1.}
    \label{tab:constraints_cosmo_params_dr1}
    \hspace{0.5cm}
    \centering
    \renewcommand{\arraystretch}{1.3}
    \setlength{\tabcolsep}{3.5pt}
    \begin{tabular}{l |c|c|c|c|c}
        \toprule\toprule
        Dataset & $\hubble$ (\kmsMpc) & \om & $\ob$ & $w_0$ & $w_a$ \\
        \midrule
        \multicolumn{6}{c}{\lcdm} \\
        \midrule
        % DESI-DR2 & -- & $0.2972^{+0.0084}_{-0.0084}$ & $> 0.0330$ & -- & -- \\
        % BAO-noDESI + DESI-DR2 & -- & $0.2969^{+0.0084}_{-0.0085}$ & $> 0.0330$ & -- & -- \\
        SN + CMB + DESI-DR2 & $68.30^{+0.28}_{-0.28}$ & $0.3025^{+0.0036}_{-0.0036}$ & $0.0483^{+0.0003}_{-0.0003}$ & -- & -- \\
        BAO-noDESI + SN + CMB + DESI-DR2 & $68.32^{+0.28}_{-0.28}$ & $0.3023^{+0.0035}_{-0.0035}$ & $0.0482^{+0.0003}_{-0.0003}$ & -- & -- \\
        \midrule
        \multicolumn{6}{c}{\wacdm} \\
        \midrule
        % DESI-DR2 & -- & $0.3397^{+0.0319}_{-0.0143}$ & $> 0.0362$ & $> -0.977$ & $< 0.336$ \\
        % BAO-noDESI + DESI-DR2 & -- & $0.3395^{+0.0323}_{-0.0141}$ & $> 0.0364$ & $> -0.978$ & $< 0.332$ \\
        SN + CMB + DESI-DR2 & $66.86^{+0.55}_{-0.55}$ & $0.3186^{+0.0056}_{-0.0055}$ & $0.0502^{+0.0008}_{-0.0008}$ & $-0.753^{+0.057}_{-0.057}$ & $-0.84^{+0.24}_{-0.22}$ \\
        BAO-noDESI + SN + CMB + DESI-DR2 & $66.88^{+0.56}_{-0.56}$ & $0.3183^{+0.0056}_{-0.0056}$ & $0.0502^{+0.0009}_{-0.0009}$ & $-0.752^{+0.057}_{-0.057}$ & $-0.85^{+0.25}_{-0.21}$ \\
        \bottomrule\bottomrule
    \end{tabular}
    % \caption{68\% credible region (1$\sigma$) or 95\% upper/lower limits of cosmological parameters under \lcdm and \wacdm for different data combinations including DESI DR2.}
    \caption{68\% credible region (1$\sigma$) of cosmological parameters under \lcdm and \wacdm for different data combinations including DESI DR2.}
    \label{tab:constraints_cosmo_params_dr2}
\end{table*}

In \lcdm, adding BAO-noDESI to the combination SN+CMB+DESI-DR1 slightly decreases the central value of \om and slightly increases the one of \hubble, while \ob remains essentially unchanged. However, we get slightly better constraints in these three parameters. We must note that some care is required when interpreting these datasets jointly within \lcdm, as the SN and CMB measurements already exhibit a $1.7\sigma$ tension between them (see Table IV of \cite{abbott2025dark}). A comparable discrepancy is found between DESI BAO and the CMB, and an even larger one between DESI BAO and SN. These tensions between datasets are what led us to consider alternative models to \lcdm in \cite{abbott2025dark}, where we found that \wacdm was the only one reconciling all our observables.

In \wacdm, the inclusion of BAO-noDESI to the combination SN+CMB+DESI-DR1 also produces minor but interesting shifts in $w_0$ and $w_a$, although, unlike in \lcdm, the constraining power remains the same. Adding BAO-noDESI slightly shifts $w_0$ toward higher values and $w_a$ toward lower values, i.e., further away from \lcdm.

In the left panel of \autoref{fig:w0wa}, we show the constraints in the $(w_0, w_a)$ plane for the different dataset combinations included in \autoref{tab:constraints_cosmo_params_dr1}. As we already concluded from \autoref{tab:constraints_cosmo_params_dr1}, the addition of DES BAO-noDESI to SN+CMB+DESI-DR1 does not tighten the constraints in this parameter space. However, it shifts the contours slightly away from \lcdm, favoring marginally higher values of $w_0$ and slightly lower values of $w_a$. For comparison purposes, we also show our best constraint from \cite{abbott2025dark} (BAO-full+SN+CMB).

In \autoref{tab:chi2_summary}, we display the $\Delta\chi^2$ between \lcdm and \wacdm for the different data combinations and their conversion to $N_{\sigma}$, following the methodology described in \autoref{sec:quantifying}. The combination SN+CMB+DESI-DR1 gives $\Delta\chi^2=-16.9$, corresponding to a $3.7\sigma$ preference for \wacdm.\footnote{This result can be compared with the one quoted by the DESI collaboration in \cite{desi2024iv}, where they found a value of $3.9\sigma$ for the same combination of datasets (but including CMB lensing).} Adding BAO-noDESI on top of this slightly increases the preference to $\Delta \chi^2=-17.6$ ($3.8\sigma$), indicating that the BAO-noDESI measurements contribute additional, albeit modest, information that strengthens the evidence for \wacdm.\footnote{This is not due to our DES BAO measurement being inconsistent with \lcdm, but rather to the existing tensions between these datasets when interpreted under \lcdm.} This increase in the significance is consistent with $w_0$ moving toward higher values and $w_a$ moving toward lower values when including BAO-noDESI, which we found in \autoref{tab:constraints_cosmo_params_dr1} and \autoref{fig:w0wa} (left panel). For comparison purposes, in \autoref{tab:chi2_summary} we also include the results from BAO-full+SN+CMB, our best constraint from \cite{abbott2025dark}, which exhibits a tension of $3.0\sigma$ with \lcdm.

Finally, ignoring correlations and combining SN+CMB+DESI-DR1 with DES BAO-full instead of BAO-noDESI, the $\Delta\chi^2$ further decreases to $\Delta\chi^2=-18.9$ ($4.0\sigma$), consistent with the results derived in \cite{abbott2025dark} for the same combination of datasets (where we used \polychord \cite{handley2015polychord} instead of \nautilus as the sampler). We emphasize that the actual value for $N_\sigma$ should be interpreted as lying between the two limiting cases: (1) removing all the overlap with DESI (i.e., using BAO-noDESI) and (2) ignoring the correlation between DES and DESI (i.e., using BAO-full). Some non-zero correlation is expected between the DES and DESI BAO measurements, but certainly not a 100\% correlation, so the true result should fall between these two extremes.

\subsubsection{Combination with DESI DR2}\label{sec:combination_dr2}

In \autoref{tab:constraints_cosmo_params_dr2}, we display the constraints on cosmological parameters, in both \lcdm and \wacdm, for various combinations of datasets that include DESI DR2.

In \lcdm, a similar trend is observed in the shifts of the parameters when adding BAO-noDESI to the combination SN+CMB+DESI-DR2 as we found when using DESI DR1, but with no improvement in the constraining power, indicating that the DESI DR2 dataset already provides tight constraints, and the inclusion of BAO-noDESI adds only modest complementary information. Again, DESI DR2 BAO and CMB results are in tension when interpreted in \lcdm, as reported in \cite{desidr2bao}.\footnote{Although the DESI analysis in \cite{desidr2bao} includes CMB lensing---unlike the present work---the reported tensions persist when CMB lensing is excluded.}

In \wacdm, just like in the case of DESI-DR1, adding BAO-noDESI to the combination SN+CMB+DESI-DR2 slightly moves $w_0$ toward higher values and $w_a$ toward lower values, i.e., further away from \lcdm. This effect was more noticeable when using DESI DR1 due to its weaker constraining power compared to DR2. Overall, these results indicate that BAO-noDESI can slightly refine dark energy constraints, but the combination of SN, CMB, and DESI data—especially DR2—dominates the parameter determination, and the fits remain robust with respect to the addition of our DES BAO dataset.

\begin{figure*}
    \centering
    \includegraphics[width=0.34\linewidth]{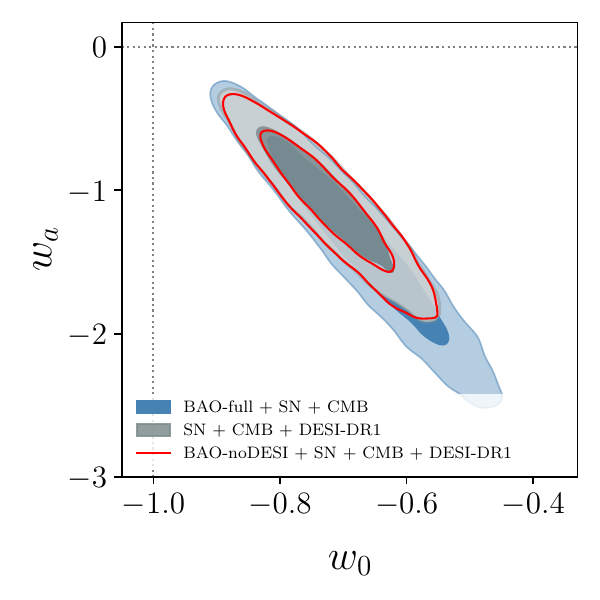}\hspace{-0.3cm}
    \includegraphics[width=0.34\linewidth]{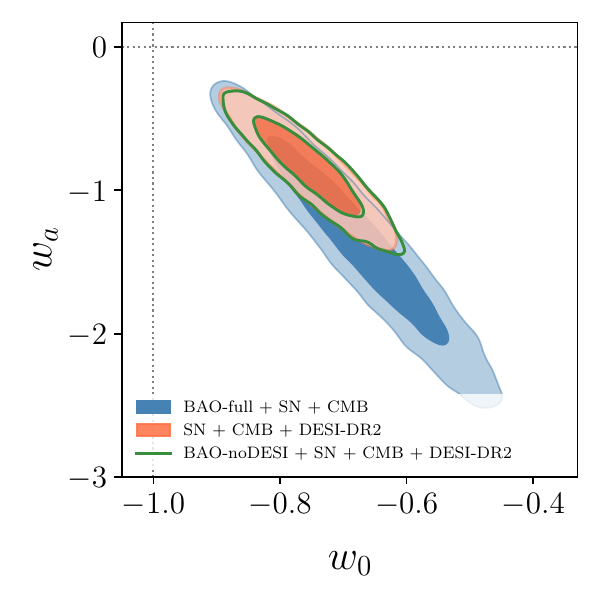}\hspace{-0.3cm}
    \includegraphics[width=0.34\linewidth]{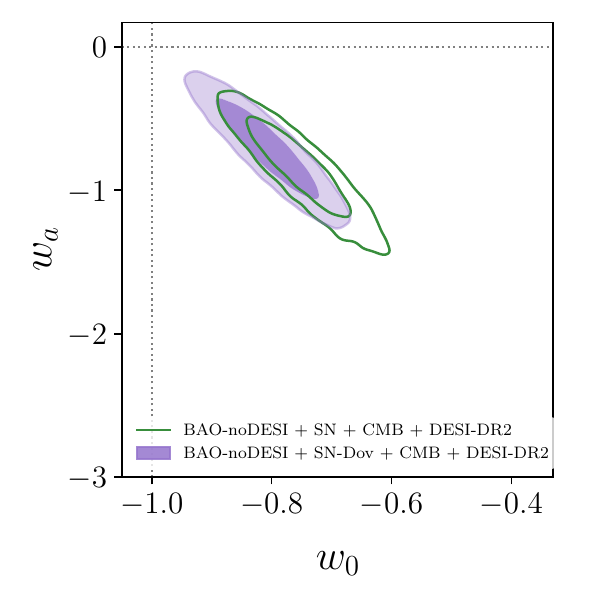}
    \caption{Constraints on the $(w_0, w_a)$ plane from different data combinations. The left and center panels show the constraints from SN+CMB+DESI and BAO-noDESI+SN+CMB+DESI, using DESI DR1 (left) and DESI DR2 (center). For comparison, we also include the BAO-full+SN+CMB constraints from \cite{abbott2025dark} in both. The right panel shows the shift in the contours (for the combination with DESI DR2) when using the DES SN–Dovekie sample.}
    \label{fig:w0wa}
\end{figure*}

In the central panel of \autoref{fig:w0wa}, we present the constraints in the $(w_0, w_a)$ plane for the different dataset combinations included in \autoref{tab:constraints_cosmo_params_dr2}. The conclusions are similar to those when using DESI DR1, although the shift was more visually apparent when using DESI DR1 (left panel) because of its weaker constraining power, which made the relative effect of BAO-noDESI more noticeable. For comparison purposes, we also show our best constraint from \cite{abbott2025dark} (BAO-full+SN+CMB). Relative to the updated combination BAO-noDESI+SN+CMB+DESI-DR2 (central panel), the constraints from \cite{abbott2025dark} are significantly weaker, illustrating the substantial improvement obtained when incorporating DESI DR2 BAO measurements.

The $\Delta\chi^2$ between the best-fit $\chi^2$ for \lcdm and \wacdm for the different data combinations and their conversion to $N_{\sigma}$ are listed in \autoref{tab:chi2_summary}. For DESI DR2, the trend is similar but the overall preference is slightly stronger. SN+CMB+DESI-DR2 alone yields $\Delta \chi^2=-19.7$ ($4.0\sigma$),\footnote{This result can be compared with the one quoted by the DESI collaboration in \cite{desidr2bao}, where they found a value of $4.2\sigma$ for the same combination of datasets (but including CMB lensing and using CamSpec \cite{efstathiou2021detailed,rosenberg2022cmb} as the Planck likelihood, which is built on {\tt NPIPE} PR4 data release from the Planck collaboration, instead of our {\tt Plik} choice based on PR3).} and the addition of BAO-noDESI shifts this marginally to $-20.0$ ($4.1\sigma$). Ignoring correlations between DES and DESI and using the full DES BAO likelihood instead gives $\Delta \chi^2=-20.5$, maintaining the $4.1\sigma$ preference. 

The combination SN+CMB+DESI-DR2 provides tighter constraints and a slightly higher $N_\sigma$ than SN+CMB+DESI-DR1, as expected from the improved precision of the DR2 dataset. In both cases, adding BAO-noDESI leads to only modest decreases in $\Delta\chi^2$, although the impact is somewhat larger for DESI DR1. Specifically, the inclusion of BAO-noDESI decreases $\Delta\chi^2$ by 0.7 for SN+CMB+DESI-DR1 (from -16.9 to -17.6), compared to 0.3 for SN+CMB+DESI-DR2 (from -19.7 to -20.0), reflecting the higher constraining power of the latter. Overall, these results show that BAO-noDESI contributes complementary information that slightly strengthens the evidence for \wacdm.

Finally, ignoring correlations and combining SN+CMB+DESI-DR2 with DES BAO-full instead of BAO-noDESI, the $\Delta\chi^2$ further decreases to $\Delta\chi^2=-20.5$ ($4.1\sigma$).

\begin{table*}[]
    \centering
    \renewcommand{\arraystretch}{1.3}
    \setlength{\tabcolsep}{4pt}
    \begin{tabular}{l|cc}
        \toprule\toprule
        Dataset & $\Delta \chi^2$ & Significance ($N_\sigma$) \\
        \midrule
        \multicolumn{3}{c}{\textbf{excluding DESI}} \\
        \midrule
        BAO-full + SN + CMB & -12.1 & 3.0 \\
        \midrule
        \multicolumn{3}{c}{\textbf{including DESI DR1}} \\
        \midrule
        % DESI-DR1 & -3.23 & 1.3 \\
        % BAO-noDESI + DESI-DR1 & -3.27 & 1.3 \\
        % BAO-full + DESI-DR1 & -3.36 & 1.3 \\
        % CMB + DESI-DR1 & -7.73 & 2.3 \\
        SN + CMB + DESI-DR1 & -16.9 & 3.7 \\
        % BAO-noDESI + CMB + DESI-DR1 & -8.21 & 2.4 \\
        BAO-noDESI + SN + CMB + DESI-DR1 & -17.6 & 3.8 \\
        % BAO-full + CMB + DESI-DR1 & -9.13 & 2.6 \\
        BAO-full + SN + CMB + DESI-DR1 & -18.9 & 4.0 \\
        \midrule
        \multicolumn{3}{c}{\textbf{including DESI DR2}} \\
        \midrule
        % DESI-DR2 & -4.62 & 1.7 \\
        % BAO-noDESI + DESI-DR2 & -4.67 & 1.7 \\
        % BAO-full + DESI-DR2 & -4.77 & 1.7 \\
        % CMB + DESI-DR2 & -6.34 & 2.0 \\
        SN + CMB + DESI-DR2 & -19.7 & 4.0 \\
        % BAO-noDESI + CMB + DESI-DR2 & -6.71 & 2.1 \\
        BAO-noDESI + SN + CMB + DESI-DR2 & -20.0 & 4.1 \\
        % BAO-full + CMB + DESI-DR2 & -7.53 & 2.3 \\
        BAO-full + SN + CMB + DESI-DR2 & -20.5 & 4.1 \\
        \bottomrule\bottomrule
    \end{tabular}
    \caption{Significance of the preference for \wacdm over \lcdm. The second column displays the $\Delta\chi^2$ between the best-fit \wacdm and \lcdm models. Negative $\Delta\chi^2$ values indicate that the \wacdm model provides a better fit. The third column shows the corresponding significances (in $N_\sigma$).}
    \label{tab:chi2_summary}
\end{table*}

\subsubsection{Impact of DES SN-Dovekie}\label{sec:dovekie}

In the right panel of \autoref{fig:w0wa}, we show the constraints in the $(w_0, w_a)$ plane when using SN-Dovekie (purple) compared to the baseline DES Y5 SN (green). We see that the contours shift toward \lcdm.

\autoref{tab:chi2_summary_Dovekie} shows the analogous results to \autoref{tab:chi2_summary} ($\Delta\chi^2$ and $N_{\sigma}$) but using SN-Dovekie instead of DES Y5 SN. Compared to the results shown in \autoref{tab:chi2_summary}, we see that using SN-Dovekie significantly alleviates the tension with \lcdm, consistent with the visual inspection of the right panel of \autoref{fig:w0wa}. For the combinations with DESI DR1, $N_\sigma$ decreases by about $0.8\sigma$, and for the combinations with DESI DR2, by about $0.9\sigma$. 

These shifts toward \lcdm are related to the fact that, in \lcdm, SN-Dovekie prefers a value of \om of $0.330\pm0.015$, significantly lower than the one preferred by the baseline DES Y5 SN ($0.352\pm0.017$), making it more consistent with the value preferred by BAO and CMB. This has an effect in the $(\om,w_0,w_a)$ space when interpreting the data under \wacdm, in particular pushing \om half a sigma lower when using SN-Dovekie, and $w_0$ and $w_a$ half a sigma closer to \lcdm \cite{popovic2025dark}. However, we still see that including DES BAO-noDESI to the combination SN-Dovekie+CMB+DESI slightly decreases $\Delta\chi^2$ and, therefore, slightly increases $N_\sigma$.

\begin{table*}[]
    \centering
    \renewcommand{\arraystretch}{1.3}
    \setlength{\tabcolsep}{4pt}
    \begin{tabular}{l|cc}
        \toprule\toprule
        Dataset & $\Delta \chi^2$ & Significance ($N_\sigma$) \\
        \midrule
        \multicolumn{3}{c}{\textbf{excluding DESI}} \\
        \midrule
        BAO-full + SN-Dovekie + CMB & -8.0 & 2.4 \\
        \midrule
        \multicolumn{3}{c}{\textbf{including DESI DR1}} \\
        \midrule
        SN-Dovekie + CMB + DESI-DR1 & -11.1 & 2.9 \\
        BAO-noDESI + SN-Dovekie + CMB + DESI-DR1 & -11.7 & 3.0 \\
        BAO-full + SN-Dovekie + CMB + DESI-DR1 & -12.7 & 3.1 \\
        \midrule
        \multicolumn{3}{c}{\textbf{including DESI DR2}} \\
        \midrule
        SN-Dovekie + CMB + DESI-DR2 & -12.4 & 3.1 \\
        BAO-noDESI + SN-Dovekie + CMB + DESI-DR2 & -12.6 & 3.1 \\
        BAO-full + SN-Dovekie + CMB + DESI-DR2 & -13.2 & 3.2 \\
        \bottomrule\bottomrule
    \end{tabular}
    \caption{Same as \autoref{tab:chi2_summary} but using DES SN-Dovekie as the SN dataset.}
    \label{tab:chi2_summary_Dovekie}
\end{table*}

\section{Conclusions}\label{sec:conclusions}

In the first part of this work, we constructed an alternative DES Y6 BAO likelihood that is fully independent of DESI, enabling robust joint analyses of the two data sets. Starting from the fiducial DES Y6 BAO sample \citep{mena2024dark}, we developed mask-level procedures to remove all regions overlapping with DESI. We defined two prescriptions for this separation: the \texttt{DR1tiles} split, optimized for combinations with DESI DR1 and DR2, and the \texttt{Deccut} split, designed for future combinations with DESI DR3 and DR4. These selections produce complementary DES no-DESI and DES DESI regions with well-defined areas and redshift distributions identical to those of the full sample. Unlike the fiducial DES Y6 BAO analysis \cite{y6-baokp}, which combined ACF, angular power spectrum (APS) and projected correlation function (PCF), here we derived the BAO signal exclusively from the ACF.

We applied the same systematics-mitigation and clustering-measurement pipelines used in the fiducial DES Y6 BAO analysis. Validation was performed with the 1,952 DES Y6 \texttt{ICE-COLA} mocks, to which we applied the same mask splits, enabling direct, one-to-one comparisons between data and simulations. Together, these elements yielded a validated and fully DESI-independent DES Y6 BAO likelihood, suitable for combination with current and upcoming DESI releases.

Although less constraining than the full sample, the no-DESI selections retained sufficient statistical power for precise BAO measurements, and their angular correlation functions remained consistent with those of the complete DES Y6 sample. For the \tilesnodesi sample, we obtained a 3.1\% BAO measurement at redshift 0.851, while the \decbelow sample yields 3.2\% (larger due to its reduced area), compared to 2.4\% for the full sample. The latter was found to be fully consistent with the ACF-only measurement reported in \cite{y6-baokp} (differences below $0.1\sigma$), as both analyses relied on the same pipeline with modest refinements.

The recovered BAO-scaling parameter, $\alpha$, showed shifts of 1.56\% (\tilesnodesi) and 2.92\% (\decbelow) relative to the full-sample result, both of which were within the statistical expectations from the mocks. The increases in $\sigma_\alpha$ after removing the DESI overlap were likewise consistent with the simulations and with the scaling expected from the reduced areas.

In the second part of the paper, we assessed the information provided by our new DES BAO-noDESI likelihood for \wacdm constraints when combined with DES SN, Planck CMB, and DESI BAO. We focused on the DES BAO \tilesnodesi sample here, which is appropriate for combinations with DESI DR1 or DR2; for future DESI releases (DR3 and DR4), the \decbelow sample should be used instead.

The combination of SN, CMB and DESI BAO data showed a preference for \wacdm over \lcdm. Using DESI DR1 yielded $\Delta\chi^2 = -16.9$ (a $3.7\sigma$ preference), which slightly strengthened to $\Delta\chi^2 = -17.6$ ($3.8\sigma$) when adding DES BAO-noDESI. Using DESI DR2 instead, the preference increased to $\Delta\chi^2 = -19.7$ ($4.0\sigma$) and $\Delta\chi^2 = -20.0$ ($4.1\sigma$), respectively. These results demonstrated that the DES BAO-noDESI measurements made a slight contribution to the preference for \wacdm when interpreting all datasets together. When ignoring correlations between DES and DESI and instead combining SN+CMB+DESI with DES BAO-full, the $\Delta\chi^2$ values decreased further: from $-17.6$ to $-18.9$ when using DESI DR1, and from $-20.0$ to $-20.5$ when using DESI DR2. The larger shifts in $\Delta\chi^2$ observed for DESI DR1 reflected its lower constraining power compared to DR2.

We also studied the impact of using the newly calibrated DES SN-Dovekie sample \cite{popovic2025dark} instead of the baseline DES Y5 SN. For the combinations with DESI DR1, the values for $N_\sigma$ mentioned before decreased by about $0.8\sigma$, and for the combinations with DESI DR2, by about $0.9\sigma$. The level of tension was, therefore, reduced when using SN-Dovekie. However, it was still above $3.0\sigma$ for all data combinations that included DES BAO-noDESI.

The BAO-fitting pipeline used in this analysis, \baofitwtheta, has been publicly released on \github at \url{https://github.com/juanejo95/BAOfit_wtheta} alongside this paper. It corresponds to a lightly updated version of the code employed in the fiducial DES Y6 BAO ACF analysis \cite{y6-baokp}. A detailed description of the code’s functionality is provided in \appendixcite{app:baofitwtheta}. Finally, we are also publicly releasing the two DESI-independent DES BAO likelihoods produced in this work (\tilesnodesi and \decbelow), which will be made available in \cosmosis. We emphasize, however, that analyses using DES BAO data should rely on the fiducial likelihood of \cite{y6-baokp} unless they are explicitly designed to be combined with DESI.

\section*{Data availability}

The two DESI-independent DES BAO likelihoods are publicly available at \url{https://github.com/cosmosis-developers/cosmosis-standard-library/tree/main/likelihood/bao/des/y6-no-desi}. 

\begin{acknowledgments}
\textit{Author's contributions:} We would like to acknowledge everyone who made this work possible. 
J. Mena-Fern\'andez divided the DES Y6 BAO mask into the DES no-DESI and DES DESI regions; applied these masks to the DES Y6 BAO sample and the COLA mocks; measured the angular two-point correlation functions; performed the BAO fits; ran the cosmological parameter inference and posterior maximization; and produced the corresponding tables and plots.
S. Avila coordinated the development of this analysis. 
A. Porredon produced the analytical covariance matrices used for the BAO fits.
H. Camacho measured the angular power spectrum from the DES no-DESI and DES DESI masks and served as an internal reviewer. 
J. Muir provided useful discussions on the posterior maximization procedure.
E. Sanchez served as an internal reviewer.
M. Vincenzi, T. Davis, R. Camilleri, P. Shah, A. Fert{\'e}, G. Campailla, M. Raveri and N. Deiosso contributed as part of the core team of the DES BAO+SN analysis, precursor of this paper.
C. Doux provided useful discussions on the tension metrics.
M. Adamow, K. Bechtol, A. Drlica-Wagner, R. A. Gruendl, W. G. Hartley, A. Pieres, E. S. Rykoff, I. Sevilla-Noarbe, E. Sheldon and B. Yanny contributed to the creation of the Y6 Gold catalog.
We would also like to acknowledge A. Raichoor for providing feedback on how to produce the DESI mask from the public tiles file, and A. J. Ross for useful discussions on the declination cut to remove the overlap with DESI DR4.

Funding for the DES Projects has been provided by the U.S. Department of Energy, the U.S. National Science Foundation, the Ministry of Science and Education of Spain, 
the Science and Technology Facilities Council of the United Kingdom, the Higher Education Funding Council for England, the National Center for Supercomputing 
Applications at the University of Illinois at Urbana-Champaign, the Kavli Institute of Cosmological Physics at the University of Chicago, 
the Center for Cosmology and Astro-Particle Physics at the Ohio State University,
the Mitchell Institute for Fundamental Physics and Astronomy at Texas A\&M University, Financiadora de Estudos e Projetos, 
Funda{\c c}{\~a}o Carlos Chagas Filho de Amparo {\`a} Pesquisa do Estado do Rio de Janeiro, Conselho Nacional de Desenvolvimento Cient{\'i}fico e Tecnol{\'o}gico and 
the Minist{\'e}rio da Ci{\^e}ncia, Tecnologia e Inova{\c c}{\~a}o, the Deutsche Forschungsgemeinschaft and the Collaborating Institutions in the Dark Energy Survey. 

The Collaborating Institutions are Argonne National Laboratory, the University of California at Santa Cruz, the University of Cambridge, Centro de Investigaciones Energ{\'e}ticas, 
Medioambientales y Tecnol{\'o}gicas-Madrid, the University of Chicago, University College London, the DES-Brazil Consortium, the University of Edinburgh, 
the Eidgen{\"o}ssische Technische Hochschule (ETH) Z{\"u}rich, 
Fermi National Accelerator Laboratory, the University of Illinois at Urbana-Champaign, the Institut de Ci{\`e}ncies de l'Espai (IEEC/CSIC), 
the Institut de F{\'i}sica d'Altes Energies, Lawrence Berkeley National Laboratory, the Ludwig-Maximilians Universit{\"a}t M{\"u}nchen and the associated Excellence Cluster Universe, 
the University of Michigan, NSF NOIRLab, the University of Nottingham, The Ohio State University, the University of Pennsylvania, the University of Portsmouth, 
SLAC National Accelerator Laboratory, Stanford University, the University of Sussex, Texas A\&M University, and the OzDES Membership Consortium.

Based in part on observations at NSF Cerro Tololo Inter-American Observatory at NSF NOIRLab (NOIRLab Prop. ID 2012B-0001; PI: J. Frieman), which is managed by the Association of Universities for Research in Astronomy (AURA) under a cooperative agreement with the National Science Foundation.

The DES data management system is supported by the National Science Foundation under Grant Numbers AST-1138766 and AST-1536171.
Data access is enabled by Jetstream2 and OSN at Indiana University through allocation PHY240006: Dark Energy Survey from the Advanced Cyberinfrastructure Coordination Ecosystem: Services and Support (ACCESS) program, which is supported by U.S. National Science Foundation grants 2138259, 2138286, 2138307, 2137603, and 2138296.
The DES participants from Spanish institutions are partially supported by MICINN under grants PID2021-123012, PID2021-128989 PID2022-141079, SEV-2016-0588, CEX2020-001058-M and CEX2020-001007-S, some of which include ERDF funds from the European Union. IFAE is partially funded by the CERCA program of the Generalitat de Catalunya.

We  acknowledge support from the Brazilian Instituto Nacional de Ci\^encia
e Tecnologia (INCT) do e-Universo (CNPq grant 465376/2014-2).

This document was prepared by the DES Collaboration using the resources of the Fermi National Accelerator Laboratory (Fermilab), a U.S. Department of Energy, Office of Science, Office of High Energy Physics HEP User Facility. Fermilab is managed by Fermi Forward Discovery Group, LLC, acting under Contract No. 89243024CSC000002.
\end{acknowledgments}

\appendix

\section{\baofitwtheta}\label{app:baofitwtheta}

\baofitwtheta (\url{https://github.com/juanejo95/BAOfit_wtheta}) is organized into four main modules, each implementing a specific set of functionalities: \texttt{utils\_data.py}, \texttt{utils\_cosmology.py}, \texttt{utils\_template.py} and \texttt{utils\_baofit.py}. Each module contains one or more classes designed to handle specific tasks, summarized below. Notebooks illustrating how to compute and plot the theoretical angular correlation function, as well as how to run BAO fits and analyze the results, are provided in the \texttt{notebooks} subfolder of the \github link.

\subsection*{utils\_data.py}
This module provides classes for handling the redshift distributions, data $w(\theta)$ and covariance used for the fits.

\begin{itemize}
    \item \texttt{RedshiftDistributions}: Loads and manages the redshift distributions.
    
    \item \texttt{GetThetaLimits}: Computes $\theta$ limits for each redshift bin to be used for the fits.
    
    \item \texttt{WThetaDataCovariance}: Handles the loading and processing of the data $w(\theta)$ and the associated covariance matrix.
\end{itemize}

\subsection*{utils\_cosmology.py}
This module handles cosmological parameters.

\begin{itemize}
    \item \texttt{CosmologicalParameters}: Initializes the cosmological parameters for a specified template cosmology.
\end{itemize}

\subsection*{utils\_template.py}
This module provides classes to initialize and compute the theoretical $w(\theta)$ that are used as templates for the BAO fits.

\begin{itemize}
    \item \texttt{TemplateInitializer}: Sets up the template $w(\theta)$ environment.

    \item \texttt{PowerSpectrumMultipoles}: Computes the multipoles of the power spectrum $P_\ell(k)$ for each redshift bin from the theoretical power spectrum $P(k)$.

    \item \texttt{CorrelationFunctionMultipoles}: Computes the multipoles of the correlation function $\xi_\ell(r)$ from the power spectrum multipoles $P_\ell(k)$.

    \item \texttt{WThetaCalculator}: Computes the angular correlation function $w(\theta)$ by projecting in redshift the correlation function multipoles $\xi_\ell(r)$ using the redshift distributions.

    \item \texttt{CellCalculator}: Computes the angular power spectra $C_\ell$ from the angular correlation function $w(\theta)$ for each redshift bin.
\end{itemize}

\subsection*{utils\_baofit.py}
This module provides classes to set up and perform BAO fits using the angular correlation function $w(\theta)$.

\begin{itemize}
    \item \texttt{WThetaModelGalaxyBias}: Initializes the model $w(\theta)$ for the linear galaxy bias fit (optional).

    \item \texttt{WThetaModelBAO}: Initializes the model $w(\theta)$ for the BAO fits.

    \item \texttt{BAOFitInitializer}: Sets up the BAO fitting environment.

    \item \texttt{BAOFit}: Performs the BAO fits and saves the results.
\end{itemize}

\bibliography{bao_DES+DESI}

\end{document}